\documentclass[9pt,numbers,legacycopyright]{sigplanconf}
\usepackage{hyperref}
\usepackage{upquote}
\usepackage{color,soul}
\usepackage{graphicx}
\usepackage{multirow}
\usepackage{booktabs}
\usepackage{dcolumn}
\usepackage{languages}
\usepackage{adjustbox}
\usepackage{comment}
\usepackage{amssymb}
\usepackage{amsmath}
\usepackage{listings,multicol}
\usepackage[figure, figure*]{hypcap}
\usepackage[labelfont=bf]{caption}
\usepackage{subcaption}
\usepackage{balance}

\newcommand{\sv}[1]{\lstinline|#1|}
\newtheorem{theorem}{Theorem}

\definecolor{amber}{rgb}{1.0, 0.75, 0.0}
\definecolor{applegreen}{rgb}{0.55, 0.71, 0.0}
\definecolor{citrine}{rgb}{0.89, 0.82, 0.04}

\newcommand\aggelos[1]{\sethlcolor{applegreen}\noindent\hl{\textit{Aggelos: #1}}}

\newcommand\strymonas{\textbf{\textit{strymonas}}}

\newenvironment{bullets}%
{\begin{list}{$\bullet$}{\setlength{\leftmargin}{1.5ex}%
      \setlength{\itemindent}{.5ex}%
      \setlength\itemsep{0.2mm}}%
\setlength{\parindent}{2ex}%
}%
{\end{list}}

\lstdefinestyle{sOcaml}{
  	language=[Objective]Caml,
  	tabsize=3,
  	keepspaces=true,
  	upquote=true,
  	alsoletter={'},
  	columns=[l]flexible,
  	flexiblecolumns=true,
  	xleftmargin=1em,
  	mathescape=true,
  	escapeinside=\`\`,
  	showstringspaces=false,
  	basicstyle={\small\ttfamily},
  	literate={+}{{$+\:$}}1
           {/}{{$/$}}1
           {>}{{$>$}}1 {<}{{$<$}}1
           {<>}{$\not=\ $}1
           {->}{{$\rightarrow$}}2
           {>=}{{$\geq$}}2 {<-}{{$\leftarrow$}}2
           {<=}{{$\leq$}}2
           {.<}{$.\langle$}1
           {>.}{$\rangle.$}1
           {.~}{$\sim$}2
           {|}{{$\mid$}}1
           {|>}{{$\triangleright$}}2
           {'a}{$\alpha$}1
           {'b}{$\beta$}1
           {'c}{$\gamma$}1
           {'s}{$\sigma$}1
           {'w}{$\omega$}1
           {'z}{$\zeta$}1
           {list}{list}1
           {forall}{$\forall\ignorespaces$}2
           {exists}{$\exists\ignorespaces$}1
           {\#\#\#}{{$\leadsto$}}3
           {=def=}{{$\stackrel{def}{=}$}}3
}

\lstnewenvironment{codeE}{\lstset{basicstyle={\ttfamily\footnotesize}}}{}
\lstnewenvironment{code}{\vspace{-1mm}}{}

\lstMakeShortInline[columns=fullflexible]|

\usepackage[firstpage]{draftwatermark}
\SetWatermarkText{\hspace*{8in}\raisebox{3.7in}%
  {\includegraphics[scale=0.1]{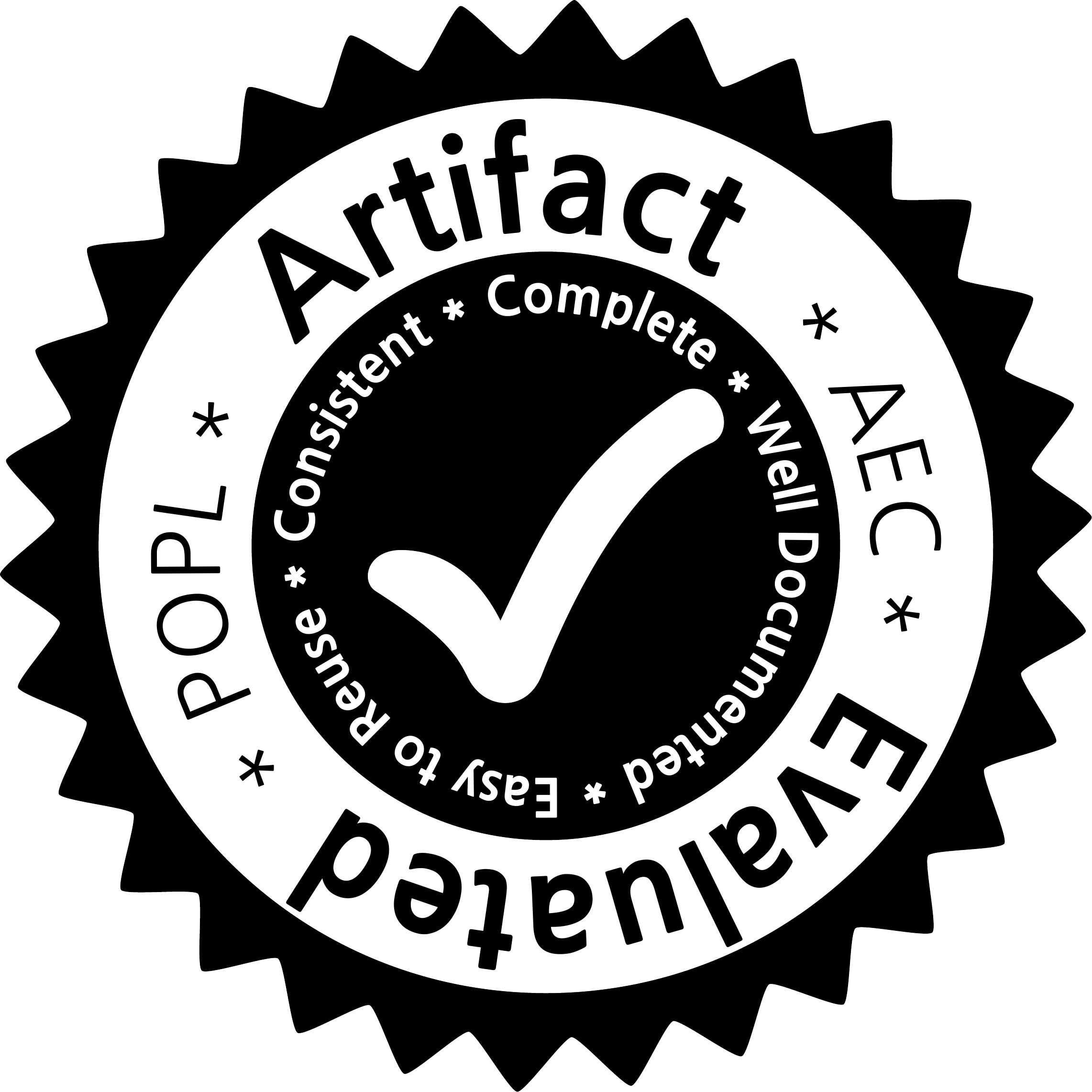}}}
\SetWatermarkAngle{0}

\begin{document}

\setlength{\pdfpageheight}{\paperheight}
\setlength{\pdfpagewidth}{\paperwidth}

\publicationrights{licensed}
\conferenceinfo{POPL'17} {January 15–-21, 2017, Paris, France.}
\copyrightyear{2017}
\copyrightdata{978-1-4503-4660-3/17/01}
\copyrightdoi{3009837.3009880}





\addtolength{\textfloatsep}{-3mm}
\addtolength{\floatsep}{-2mm}
\addtolength{\intextsep}{-3mm}
\addtolength{\abovecaptionskip}{-3mm}
\addtolength{\belowcaptionskip}{-2mm}

\addtolength{\dbltextfloatsep}{-1mm}


\title{Stream Fusion, to Completeness	}

\authorinfo{Oleg Kiselyov}
           {Tohoku University, Japan}
           {oleg@okmij.org}

\authorinfo{Aggelos Biboudis}
           {University of Athens, Greece}
           {biboudis@di.uoa.gr}

\authorinfo{Nick Palladinos}
           {Nessos IT S.A. Athens, Greece}
           {npal@nessos.gr}

\authorinfo{Yannis Smaragdakis}
           {University of Athens, Greece}
           {smaragd@di.uoa.gr}

\maketitle

\begin{abstract}


Stream processing is mainstream (again): Widely-used stream libraries
are now available for virtually all modern OO and functional
languages, from Java to C\# to Scala to OCaml to Haskell. Yet
expressivity and performance are still lacking. For instance, the
popular, well-optimized Java 8 streams do not support the |zip|
operator and are still an order of magnitude slower than hand-written
loops.

We present the first approach that represents the full generality of
stream processing and eliminates overheads, via the use of
staging. It is based on an unusually rich semantic model of stream
interaction. We support \emph{any} combination of
zipping, nesting (or flat-mapping),
sub-ranging, filtering, mapping---of finite or
infinite streams.  Our model captures idiosyncrasies that a programmer
uses in optimizing stream pipelines, such as rate differences and the
choice of a ``for'' vs. ``while'' loops. Our approach delivers
hand-written--like code, but automatically.  It explicitly avoids the
reliance on black-box optimizers and sufficiently-smart compilers,
offering highest, guaranteed and portable performance.

Our approach relies on high-level concepts that are then readily
mapped into an implementation. Accordingly, we have two distinct
implementations: an OCaml stream library, staged via MetaOCaml, and a
Scala library for the JVM, staged via LMS.  In both cases, we derive
libraries richer and simultaneously many tens of times faster than
past work. We greatly exceed in performance the standard stream
libraries available in Java, Scala and OCaml, including the
well-optimized Java 8 streams.


\end{abstract}




\section{Introduction}


\emph{Stream processing} defines a pipeline of operators that
transform, combine, or reduce (even to a single scalar) large amounts
of data.  Characteristically, data is accessed strictly linearly
rather than randomly and repeatedly---and processed uniformly. The
upside of the limited expressiveness is the opportunity to process
large amount of data efficiently, in constant and small
space. Functional stream libraries let us easily build such pipelines,
by composing sequences of simple transformers such as |map| or
|filter| with producers (backed by an array, a file, or a generating
function) and consumers (reducers). The purely applicative approach of
building a complex pipeline from simple immutable pieces simplifies
programming and reasoning: the assembled pipeline is an executable
specification. To be practical, however, a library has to be
efficient: at the very least, it should avoid creating intermediate
structures (files, lists, etc.) whose size grows with the length of
the stream.


Most modern programming languages---Java, \scala, \cs, \fs, OCaml, Haskell,
Clojure, to name a few---currently offer functional stream libraries. They
all provide basic mapping and filtering. Handling of infinite,
nested or parallel (zipping) streams is rare---especially all in the same
library. Although all mature libraries avoid unbounded intermediate structures,
they all suffer, in various degrees, from the overhead of abstraction and
compositionality: extra function calls, the creation of
closures, objects and other bounded intermediate structures.


An excellent example is the Java 8 Streams, often taken as the
standard of stream libraries. It stresses performance: e.g., streaming
from a known source, such as an array, amounts to an ordinary loop,
well-optimized by a Java JIT
compiler~\cite{biboudis_streams_2015}. However, Java 8 Streams are
still much slower than hand-optimized loops for non-trivial pipelines
(e.g., over 10x slower on the standard cartesian product
benchmark~\cite{biboudis_clash_2014}). Furthermore, the library cannot
handle (`zip') several streams in parallel\footnote{\relax
One could
emulate \texttt{zip} using
\texttt{iterator} from push-streams---at significant
drop in performance. }
and cannot deal with
nesting of infinite streams. These are not mere omissions: infinite
nested streams demand a different iteration model, which is hard to
efficiently implement with a simple loop.

This paper presents \strymonas{}: a streaming library design that offers both high expressivity
and \emph{guaranteed}, highest performance. First, we support the
full range of streaming operators
(a.k.a. stream \emph{transformers} or \emph{combinators}) from
past libraries: not just |map| and |filter| but also
sub-ranging (|take|), nesting (|flat_map|---a.k.a. |concatMap|) and parallel (|zip_with|)
stream processing.  All operators are freely composable: e.g.,
|zip_with| and |flat_map| can be used together, repeatedly, with
finite or infinite streams.
Our novel stream representation captures the essence of stream
processing for virtually all combinators examined in past literature.

Second, our stream representation allows eliminating the abstraction
overhead altogether, for the full set of stream operators.
We perform \emph{stream fusion} (\S\ref{sec:fusion-problem})
and other aggressive optimization. The generated code contains no extra
heap allocations in the main loop (Thm.\ref{thm:expr}). By not generating
tuples or other objects, we avoid the overhead of dynamic object
construction and pattern-matching, and also the hidden,
often significant overhead of memory pressure and boxing of primitive
types. The result not merely approaches but attains the
performance of hand-optimized code, from the simplest to the most
complex cases, up to \emph{well over} the complexity point where
hand-written code becomes infeasible. Although the library combinators
are purely functional and freely composable, the actual running stream
code is loop-based, highly tangled and imperative.

Our technique relies on staging (\S\ref{sec:msp}), a form of
metaprogramming, to achieve guaranteed stream fusion. This is in
contrast to past use of source-to-source transformations of functional
languages \cite{kelsey_realistic_1989}, of AST run-time rewriting
\cite{murray_steno:_2011,nick_palladinos_linqoptimizer:_2013},
compile-time macros~\cite{aleksandar_prokopec_scalablitz:_2013} or
Haskell GHC
\textsc{Rules}~\cite{jones_playing_2001,coutts_stream_2007} to express
domain-specific streaming optimizations. Rather than relying on an
optimizer to eliminate artifacts of stream composition, we do not
introduce the artifacts in the first place. Our library transforms
highly abstract stream pipelines to code fragments that use the most
suitable imperative features of the host language. The appeal of
staging is its certainty and guarantees. Unlike the aforementioned
techniques, staging also ensures that the generated code is well-typed and
well-scoped, by construction. We discuss the trade-offs of
staging in \S\ref{sec:why-staging}.




Our work describes a general approach, and not just a single library
design. To demonstrate the generality of the principles,
we implemented two library versions~\footnote{\url{https://strymonas.github.io/}.},
in diverse settings. The
first is an OCaml library, staged with BER MetaOCaml
\cite{kiselyov_design_2014}. The second is a Scala library (also usable by
client code in Java and other JVM languages), staged with Lightweight Modular
Staging (LMS) \cite{rompf_lms_2012}.

We evaluate \strymonas{} on a suite of benchmarks
(\S\ref{sec:experiments}), comparing with hand-written code as
well as with other stream libraries (including Java 8 Streams).
Our staged implementation is up to more than two orders-of-magnitude
faster than standard
Java/Scala/OCaml stream libraries, matching the performance of
hand-optimized loops. (Indeed, we occasionally had to improve hand-written
baseline code, because it was slower than the library.)

Thus, our contributions are: (i) the principles and the design of
stream libraries that support the widest set of operations from past
libraries and also permit the full elimination of abstraction
overhead. The main principle is a novel representation of streams
that captures rate properties of stream transformers and the form of
termination conditions, while separating and abstracting components of the
entire stream state. This decomposition of the essence of stream
iteration is what allows us to perform very aggressive optimization, via
staging, regardless of the streaming pipeline configuration.
(ii) The implementation of the design in terms of two distinct
library versions for different languages and staging methods: OCaml/MetaOCaml
and Scala/JVM/LMS.

\section{Overview: A Taste of the Library}
\label{sec:overview}

We first give an overview of our approach, presenting the client
code (i.e., how the library is used) alongside the generated code
(i.e., what our approach achieves). Although we have
implemented two separate library versions, one for OCaml and one for
Scala/JVM languages, for simplicity, all examples in the paper will
be in (Meta)OCaml, which was also our original implementation.

For the sake of exposition, we take a few liberties with the OCaml
notation, simplifying the syntax of the universal and existential
quantification and of sum data types with record components. (The
latter simplification---inline records---is supported in the latest,
4.03, version of OCaml.)  The paper is accompanied by the complete
code for the \strymonas{} library (as an open-source repository), also including
our examples, tests, and benchmarks.

MetaOCaml is a dialect of OCaml with staging annotations |.<e>.| and
|.~e|, and the |code| type
\cite{taha_gentle_2004,kiselyov_design_2014}. In the Scala version of
our library, staging annotations are implicit: they are determined by
inferred types. Staging annotations are optimization directives,
guiding the partial evaluation of library expressions. Thus, staging
annotations are not crucial to understanding what our library can express,
only how it is optimized.
On first read, staging
annotations
may be simply disregarded. We get back to them, in detail, in
\S\ref{sec:msp}.

The (Meta)OCaml library interface is given in
Figure~\ref{f:interface}. The library includes stream producers (one
generic---|unfold|, and one specifically for arrays---|of_arr|), the generic
stream consumer (or stream reducer) |fold|, and a number of stream
transformers. Ignoring |code| annotations, the signatures are
standard. For instance, the generic |unfold| combinator takes a
function from a state, |'z|, to a value |'a| and a new state
(or nothing at all), and, given an initial state |'z|, produces an
opaque stream of |'a|s.

\begin{figure}[tb!p]
\vspace{-1.8mm}
\begin{lstlisting}
$\textrm{Stream representation (abstract)}$
type 'a stream

$\textrm{Producers}$
val of_arr : 'a array code -> 'a stream
val unfold : ('z code -> ('a * 'z) option code) ->
             'z code -> 'a stream

$\textrm{Consumer}$
val fold : ('z code -> 'a code -> 'z code) ->
           'z code -> 'a stream -> 'z code

$\textrm{Transformers}$
val map        : ('a code -> 'b code) -> 'a stream ->
                 'b stream
val filter     : ('a code -> bool code) ->
                 'a stream -> 'a stream
val take       : int code -> 'a stream -> 'a stream
val flat_map   : ('a code -> 'b stream) ->
                 'a stream -> 'b stream
val zip_with   : ('a code -> 'b code -> 'c code) ->
                 ('a stream -> 'b stream -> 'c stream)
\end{lstlisting}
\caption{The library interface}
\label{f:interface}
\end{figure}

The first example is summing the squares of elements of an array
|arr|---in mathematical notation, $\sum{a_i^2}$. The
code
\vspace{-0.5mm}
\begin{lstlisting}
let sum = fold (fun z a -> .<.~a + .~z>.) .<0>.

of_arr .<arr>.
  |> map (fun x -> .<.~x * .~x>.)
  |> sum
\end{lstlisting}
is not far from the mathematical notation. Here,
$\triangleright$, like the similar operator in F\#, is the
inverse function application: argument to the left, function to the
right. The stream components are first-class and hence may be
passed around, bound to identifiers and shared; in short, we can build
libraries of more complex components.

\noindent In this simple example, the generated code is understandable:

\begin{code}
let s_1 = ref 0 in
let arr_2 = arr in
 for i_3 = 0 to Array.length arr_2 -1 do
   let el_4 = arr_2.(i_3) in
   let t_5 = el_4 * el_4 in
   s_1 := t_5 + !s_1
 done;
!s_1
\end{code}

It is relatively easy to see which part of the code came from which
part of the pipeline ``specification''. The generated code has no
closures, tuples or other heap-allocated structures: it looks as if it
were hand-written by a competent OCaml programmer. The iteration
is driven by the source operator, |of_arr|, of the pipeline. This is
precisely the iteration pattern that Java 8 streams optimize. As we will
see in later examples, this is but one of the optimal iteration patterns
arising in stream pipelines.

The next example sums only some elements:
\begin{lstlisting}
let ex = of_arr .<arr>. |> map (fun x -> .<.~x * .~x>.)

ex |> filter (fun x -> .<.~x mod 17 > 7>.) |> sum
\end{lstlisting}
We have abstracted out the
mapped stream as |ex|. The earlier example is, hence,
\lstinline{ex |> sum}. The current example applies |ex| to the more complex
summator that first filters out elements before summing the
rest. The next example limits the number of summed elements to a
user-specified value |n|
\begin{lstlisting}
ex |> filter (fun x -> .<.~x mod 17 >7>.)
   |> take .<n>.
   |> sum
\end{lstlisting}
We stress that the limit is applied to the filtered stream, not to the original
input; writing this
example in mathematical notation would be cumbersome.
The generated code
\begin{code}
let s_1 = ref 0 in
let arr_2 = arr in
let i_3 = ref 0 in
let nr_4 = ref n in
while !nr_4 > 0 && !i_3 <= Array.length arr_2 -1 do
   let el_5 = arr_2.(! i_3) in
   let t_6 = el_5 * el_5 in
   incr i_3;
   if t_6 mod 17 > 7
   then (decr nr_4; s_1 := t_6+!s_1)
done; ! s_1
\end{code}
again looks as if it were handwritten, by a competent
programmer. However, compared to the first example, the code is more tangled;
for example, the |take .<n>.| part of the pipeline contributes to
three separate places in the code: where the |nr_4| reference cell is created,
tested and mutated. The iteration pattern is more complex. Instead of
a |for| loop there is a |while|, whose termination conditions come
from two different pipeline operators: |take| and |of_arr|.

The dot-product of two arrays |arr1| and |arr2| looks just as simple
\begin{lstlisting}
zip_with (fun e1 e2 -> .<.~e1 * .~e2>.)
         (of_arr .<arr1>.)
         (of_arr .<arr2>.) |> sum
\end{lstlisting}
showing off the zipping of two streams, with the straightforward, again
hand-written quality, generated code:
\begin{code}
let s_17 = ref 0 in
let arr_18 = arr1 in let arr_19 = arr2 in
 for i_20 = 0 to
  min (Array.length arr_18 -1)
      (Array.length arr_19 -1) do
   let el_21 = arr_18.(i_20) in
   let el_22 = arr_19.(i_20) in
   s_17 := el_21 * el_22 + !s_17
 done; ! s_17
\end{code}
The optimal iteration pattern is different still (though
simple): the loop condition as well as the loop body are equally
influenced by two |of_arr| operators.

In the final, complex example we zip two complicated streams.  The
first is a finite stream from an array, mapped, subranged, filtered
and mapped again. The second is an infinite stream of natural numbers
from 1, with a filtered flattened nested substream. After zipping,
we fold everything into a list of tuples.


\definecolor{light-gray}{gray}{0.6}
\sethlcolor{light-gray} 
\begin{lstlisting}
zip_with (fun e1 e2 -> .<(.~e1,.~e2)>.)
 (of_arr .<arr1>. (* 1st stream *)
   |> map (fun x -> .<.~x * .~x>.)
   |> take .<12>.
   |> filter (fun x -> .<.~x mod 2 = 0>.)
   |> map (fun x -> .<.~x * .~x>.))
 (iota .<1>.      (* 2nd stream *)
   |> flat_map (fun x -> iota .<.~x+1>. |> take .<3>.)
   |> filter (fun x -> .<.~x mod 2 = 0>.))
 |> fold (fun z a -> .<.~a :: .~z>.) .<[]>.
\end{lstlisting}
\sethlcolor{yellow}

We did not show any types, but they exist (and have been
inferred). Therefore, an attempt to use an invalid operation on stream
elements (like concatenating integers or applying an ill-fitting
stream component) will be immediately rejected by the type-checker.

Although the above pipeline is purely functional, modular and rather
compact, the generated code (shown in Appendix
A of the extended version) is
large, entangled and highly imperative. Writing such code correctly
by hand is clearly challenging.

\section{Stream Fusion Problem}
\label{sec:fusion-problem}

The key to an expressive and performant stream library is a
representation of streams that fully captures the generality of
streaming pipelines and allows desired optimizations.  To understand
how the representation affects implementation and optimization
choices, we review past approaches. We see that, although some of them
take care of the egregious overhead, none manage to eliminate all of
it: the assembled stream pipeline remains slower than hand-written
code.

The most straightforward representation of streams is a linked list,
or a file, of elements. It is also the least performing. The first
example in \S\ref{sec:overview}, of summing squares, will entail:
(1) creating a stream from an array by copying all elements into it;
(2) traversing the list creating another stream, with squared elements;
(3) traversing the result, summing the elements. We end up creating three
intermediate lists. Although the whole processing still takes
time linear in the size of the stream, it requires repeated traversals
and the production of linear-size intermediate structures.
Also, this straightforward representation cannot cope with
sources that are always ready with an element: ``infinite streams''.

The problem, thus, is deforestation \cite{wadler-deforestation}: eliminating
intermediate, working data structures. For streams, in particular, deforestation
is typically called ``stream fusion''. One can discern two main groups of stream
representations that let us avoid building intermediate data
structures of unbounded size.

\paragraph{Push Streams.}
The first, heavily algebraic approach, represents a stream by its
reducer (the fold operation) \cite{meijer-functional}.
If we introduce the ``shape functor'' for a
stream with elements of type |'a| as
\begin{code}
type ('a,'z) stream_shape =
   | Nil
   | Cons of 'a * 'z
\end{code}
then the stream is formally defined as:\footnote{\relax
Strictly speaking, \texttt{stream} should be a record type:
in OCaml, only record or object components may have
the type with explicitly quantified type variables. For the sake of
clarity we lift this restriction in the paper.}
\begin{code}
type 'a stream = forall'w. (('a,'w) stream_shape -> 'w) -> 'w
\end{code}
A stream of |'a|s is hence a function with the ability to turn any
generic ``folder'' (i.e., a function from |('a,'w) stream_shape| to
|'w|) to a single |'w|. The ``folder'' function is formally called an
|F|-algebra for the |('a,-) stream_shape| functor.

For instance, an array is easily representable as such a fold:
\begin{code}
let of_arr : 'a array -> 'a stream =
   fun arr -> fun folder ->
   let s = ref (folder Nil) in
   for i=0 to Array.length arr - 1 do
     s := folder (Cons (arr.(i),!s))
   done; !s
\end{code}

Reducing a stream with the reducing
function |f| and the initial value |z| is especially straightforward in
this representation:
\begin{code}
let fold : ('z -> 'a -> 'z) -> 'z -> 'a stream -> 'z =
 fun f z str ->
  str (function Nil -> z | Cons (a,x) -> f x a)
\end{code}

More germane to our discussion is that mapping over the stream (as well as |filter|-ing and
|flat_map|-ing) are also easily expressible, without creating any variable-size
intermediate data structures:
\begin{code}
let map : ('a -> 'b) -> 'a stream -> 'b stream =
 fun f str ->
  fun folder -> str (fun x -> match x with
  | Nil        -> folder Nil
  | Cons (a,x) -> folder (Cons (f a,x)))
\end{code}

A stream element |a| is transformed ``on the fly'' without collecting in
working buffers. Our sample squaring-accumulating pipeline runs in
constant memory now. Deforestation, or stream fusion, has been
accomplished. The simplicity of this so-called ``push stream'' approach
makes it popular: it is used, for example, in the reducers of Clojure
as well as in the OCaml ``batteries'' library. It is also the basis
of Java 8 Streams, under an object-oriented reformulation of the same
concepts.

In push streams, it is the stream producer, e.g., |of_arr|, that
drives the optimal execution of the stream. Implementing |take| and
other such combinators that restrict the processing to a prefix of the
stream requires extending the representation with some sort of a
``feedback'' mechanism (often implemented via exceptions).  Where push
streams stumble is the zipping of two streams, i.e., the processing of
two streams in parallel. This simply cannot be done with constant
per-element processing cost. Zipping becomes especially complicated
(as we shall see in \S\ref{sec:zip}) when the two pipelines contain
nested streams and hence produce elements at generally different
rates.\footnote{The Reactive Extensions (Rx) framework~\cite{rx} gives a
  real-life example of the complexities of implementing \texttt{zip}.
  Rx is push-based and supports \texttt{zip} at the cost of
  maintaining an unbounded intermediate queue. This deals with the
  ``backpressure in Zip'' issue, extensively-discussed in
  the Rx github repo. Furthermore, Rx seems to have abandoned blocking
  zip implementations since 2014.}

\paragraph{Pull Streams.}
An alternative representation of streams, pull streams, has a long pedigree,
all the way from the
generators of Alphard \cite{alphard} in the '70s. These are objects that implement two
methods: |init| to initialize the state and obtain the first element,
and |next| to advance the stream to the next element, if any. Such
a ``generator'' (or IEnumerator, as it has come to be popularly known) can
also be understood algebraically---or rather,
co-algebraically. Whereas push streams represent a
stream as a fold, pull streams, dually, are the expression
of an \emph{unfold} \cite{meijer-functional,Gibbons-unfold}:\footnote{\relax
For the sake of explanation, we took another liberty with the
OCaml notation, avoiding the GADT syntax for the existential.}
\begin{code}
type 'a stream = exists's. 's * ('s -> ('a,'s) stream_shape)
\end{code}
The stream is, hence, a pair of the current state and the so-called
``step'' function that, given a state, reports the end-of-stream
condition |Nil|, or the current element and the next state. (Formally, the
step function is the F-co-algebra for the |('a,-) stream_shape| functor.)
The existential quantification over the state keeps it private: the only
permissible operation is to pass it to the step function.

When an array is represented as a pull stream, the state is the tuple
of the array and the current index:
\begin{code}
let of_arr : 'a array -> 'a stream =
  let step (i,arr) =
    if i < Array.length arr
       then Cons (arr.(i), (i+1,arr)) else Nil
  in fun arr -> ((0,arr),step)
\end{code}
The step function---a pure combinator rather than a
closure---dereferences the current element and advances the index. Reducing the
pull stream now requires an iteration, of repeatedly calling |step|
until it reports the end-of-stream. (Although the types of
|of_arr|, |fold|, and |map|, etc. nominally remain the same,
the meaning of |'a stream| has changed.)
\begin{code}
let fold : ('z -> 'a -> 'z) -> 'z -> 'a stream -> 'z =
 fun f z (s,step) ->
  let rec loop z s = match step s with
  | Nil        -> z
  | Cons (a,t) -> loop (f z a) t
  in loop z s
\end{code}
With pull streams, it is the reducer, i.e., the stream consumer, that drives
the processing. Mapping over the stream
\begin{code}
let map : ('a -> 'b) -> 'a stream -> 'b stream =
  fun f (s,step) ->
    let new_step = fun s -> match step s with
    | Nil        -> Nil
    | Cons (a,t) -> Cons (f a, t)
    in (s,new_step)
\end{code}
merely transforms its step function: |new_step| calls the old step and
|map|s the returned current element, passing it immediately to the
consumer, with no buffering. That is, like push streams, pull streams also
accomplish fusion.
Befitting their co-algebraic nature, pull streams can represent both
finite and infinite streams. Stream combinators, like |take|, that cut
evaluation short are also
easy. On the other hand, skipping elements (filtering) and nested
streaming is more complex with pull streams, requiring the
generalization of the |stream_shape|, as we shall see in
\S\ref{sec:implementation}.  The main advantage of pull streams over
push streams is in expressiveness: pull streams have the ability to
process streams in parallel, enabling |zip_with| as well as more complex stream
merging. Therefore, we take pull streams as the basis of our library.

\paragraph{Imperfect Deforestation.}
Both push and pull streams eliminate the intermediate lists
(variable-size buffers) that plague a naive implementation of the
stream library. Yet they do not eliminate all the abstraction overhead.
For example, the |map| stream combinator transforms
the current stream element by passing it to some function |f| received
as an argument of |map|. A hand-written implementation would have no other
function calls. However, the pull-stream
|map| combinator introduces a closure:
|new_step|, which receives a |stream_shape| value from the
old |step|, pattern-matches on it and constructs the new |stream_shape|.
The push-stream |map| has the same problem: The step function
of |of_arr| unpacks the current state and then packs the array and the
new index again into the tuple. This repeated deconstruction and
construction of tuples and co-products is the abstraction overhead,
which a complete deforestation should eliminate, but pull and push
streams, as commonly implemented, do not.
Such ``constant'' factors make
library-assembled stream processing much slower than the hand-written
version (by up to two orders of magnitude---see \S\ref{sec:experiments}).

\section{Staging Streams}

A well-known way of eliminating abstraction overhead and
delivering ``abstraction without guilt'' is program generation:
compiling a high-level abstraction into efficient code.
In fact, the original deforestation algorithm in the literature
\cite{wadler-deforestation} is closely related to partial evaluation
\cite{sorensen-unifying}. This section introduces staging: one
particular, manual technique of partial evaluation. It lets us
achieve our goal of eliminating all abstraction
overhead from the stream library. Perfect stream fusion with staging
is hard: \S\ref{sec:simple-staging} shows that straightforward staging
(or automated partial evaluation) does not achieve full deforestation.
We have to re-think general stream processing
(\S\ref{sec:overhead-elim}).

\subsection{Multi-Stage Programming}
\label{sec:msp}

Multi-stage programming (MSP), or \emph{staging} for short, is a way
to write programs that generate programs. MSP may be thought of as a
principled version of the familiar ``code templates'', where the
templates ensure by their very construction that the generated code is not only
syntactically well-formed but also well-scoped and well-typed.

In this paper we use BER MetaOCaml~\cite{kiselyov_design_2014},
which is a dialect of OCaml with MSP extensions.
The first MSP feature is brackets, \sv{.<} and \sv{>.}, which
enclose a code template. For example, \sv{.<1+2>.} is a template
for generating code to add two literals 1 and 2.
\begin{lstlisting}
let c = .<1 + 2>.
### val c : int code = .<1 + 2>.
\end{lstlisting}
The output of the interpreter demonstrates that the code template is a
first-class object; moreover, it is a value: a \emph{code value}.
MetaOCaml can print such values, and also write them into a file to
compile it later. The code value is typed: our sample template
generates integer-valued code.

As behooves templates, they can have holes to splice-in other
templates. The splicing MSP feature, |.~|, is called an
\emph{escape}. In the following example, the template |cf| has two
holes, to be filled in with the same expression. Then |cf c| fills
the holes with the expression |c| created earlier.
\begin{lstlisting}
let cf x = .<.~x + .~x>.
### val cf : int code -> int code = <fun>
cf c
### - : int code = .<(1 + 2) + (1 + 2)>.
\end{lstlisting}

One may regard brackets and escapes as annotating code: which portions should
be evaluated as usual (at the present stage, so to speak) and which in the
future (when the generated code is compiled and run).

\subsection{Simple Staging of Streams}
\label{sec:simple-staging}

We can turn a library into, effectively, a compiler of efficient
code by adding staging annotations. This is not a simple matter of
annotating one of the standard definitions (either pull- or push-style)
of |'a stream|, however. To see this, we next consider staging
a set of pull-stream
combinators. Staging helps with performance, but the abstraction overhead still remains.

The first step in using staging is the so-called ``binding-time
analysis'': finding out which values can be known only at run-time
(``dynamically'') and what is known already at code-generation time,
(``statically'') and hence can be pre-computed. Partial evaluators
perform binding-time analysis, with various degrees of sophistication
and success, automatically and opaquely. In staging, binding-time analysis is
manual and explicit.

We start with the pull streams |map| combinator, which, recall,
has a type signature:
\begin{code}
type 'a stream = exists's. 's * ('s -> ('a,'s) stream_shape)
val map : ('a -> 'b) -> 'a stream -> 'b stream
\end{code}
Its first argument, the mapping function |f|, takes the current
stream element, which is clearly
not known until the processing pipeline is run. The result is likewise
dynamic. However, the mapping operation itself can be known
statically. Hence the staged |f| may be given the type
|'a code -> 'b code|: given code to compute |'a|s, the mapping function, |f|,
is a static way to produce code to compute |'b|s.

The second argument of |map| is the pull stream,
a tuple of the
current state (|'s|) and the step function. The state is not known
statically. The result of the step function
depends on the current state and, hence, is fully dynamic. The step
function itself, however, can be statically known.
Hence
we arrive at the following type of the staged stream
\begin{code}
type 'a st_stream =
  exists's. 's code * ('s code -> ('a,'s) stream_shape code)
\end{code}
Having done such binding-time analysis for the arguments of the |map|
combinator, it is straightforward to write the staged |map|,
by annotating---i.e., placing brackets and escapes on---the original
|map| code according to the decided binding-times:
\begin{code}
let map : ('a code -> 'b code) ->
           'a st_stream -> 'b st_stream =
  fun f (s,step) ->
    let new_step = fun s -> .<match .~(step s) with
    | Nil        -> Nil
    | Cons (a,t) -> Cons (.~(f .<a>.), t)>.
    in (s,new_step)
\end{code}
The combinators |of_arr| and |fold| are staged analogously. We use
the method of \cite{inoue-reasoning} to prove the correctness, which
easily applies to this case, given that |map| is non-recursive. The sample
processing pipeline (the first example from \S\ref{sec:overview})
\begin{code}
of_arr .<[|0;1;2;3;4|]>.
       |> map (fun a -> .<.~a * .~a>.)
       |> fold (fun x y -> .<.~x + .~y>.) .<0>.
\end{code}
then produces the following code:
\begin{codeE}
- : int code = .<
let rec loop_1 z_2 s_3 =
  match match match s_3 with
        | (i_4,arr_5) ->
          if i_4 < (Array.length arr_5)
          then Cons ((arr_5.(i_4)),
                     ((i_4 + 1), arr_5))
          else Nil
    with
    | Nil  -> Nil
    | Cons (a_6,t_7) -> Cons ((a_6 * a_6), t_7)
  with
  | Nil  -> z_2
  | Cons (a_8,t_9) -> loop_1 (z_2 + a_8) t_9 in
loop_1 0 (0, [|0;1;2;3;4|])>.
\end{codeE}
As expected, no lists, buffers or other variable-size data
structures are created. Some constant overhead is gone too: the
squaring operation of |map| is inlined. However, the triple-nested
|match| betrays the remaining overhead of constructing and
deconstructing |stream_shape| values. Intuitively, the clean abstraction
of streams (encoded in the pull streams type of |'a stream|) isolates
each operator from others. The result does not take advantage of
the property that, for this pipeline (and others of the same style),
the looping of all three operators (|of_arr|, |map|, and |fold|) will
synchronize, with all of them processing elements until the same last one.
Eliminating the overhead requires a different computation model for
streams.


\section{Eliminating All Abstraction Overhead in Three Steps}
\label{sec:overhead-elim}

We next describe how to purge all of the stream library abstraction
overhead and generate code of hand-written quality and performance. We
will be continuing the simple running example of the earlier sections,
of summing up squared elements of an
array. (\S\ref{sec:implementation} will later lift the same insights
to more complex pipelines.) As in \S\ref{sec:simple-staging}, we will
be relying on staging to generate well-formed and well-typed code. The
key to eliminating abstraction overhead from the generated code is to
move it to a generator, by making the generator take better advantage
of the available static knowledge. This is easier said than done: we
have to use increasingly more sophisticated transformations of the
stream representation to expose more static information and make it
exploitable. The three transformations we show next require
more-and-more creativity and domain knowledge, and cannot be performed
by a simple tool, such as an automated partial evaluator. In the
process, we will identify three interesting concepts in stream
processing: the structure of iteration (\S\ref{sec:fusing-stepper}),
the state kept (\S\ref{sec:state-fusing}), and the optimal kind of loop
construct and its contributors (\S\ref{sec:imperative-loops}).

\subsection{Fusing the Stepper}
\label{sec:fusing-stepper}

Modularity is the cause of the abstraction overhead we observed in
\S\ref{sec:simple-staging}: structuring
the library as a collection of composable components forces them to
conform to a single interface. For example, each component has to
use the uniform stepper function interface (see the |st_stream| type)
to report the next stream element or
the end of the stream. Hence, each component has to generate code to
examine (deconstruct) and construct the |stream_shape| data type.

At first glance, nothing can be done about this: the
result of the step function, whether it is |Nil| or a |Cons|,
depends on the current state, which is surely not known until the
stream processing pipeline is run. We do know however
that the step function invariably returns either |Nil| or a
|Cons|, and the caller must be ready to handle both alternatives. We
should exploit this static knowledge.

To statically (at code generation-time) make sure that the caller of
the step function handles both alternatives of its result, we
have to change the function to accept a pair of handlers:
one for a |Nil| result and one for a |Cons|.
In other words, we have to change the result's representation,
from the sum |stream_shape| to a product of eliminators.
Such a replacement effectively removes the need to construct the
|stream_shape| data type at run-time in the first place. Essentially,
we change |step| to be in
continuation-passing style, i.e., to accept the continuation for its
result. The |stream_shape| data type nominally remains, but it becomes the
argument to the continuation and we mark its variants as
statically known (with no need to construct it at
run-time). All in all, we arrive at the following type for
the staged stream
\begin{code}
type 'a st_stream =
 exists's. 's code *
   (forall'w. 's code ->
     (('a code,'s code) stream_shape -> 'w code) ->
         'w code)
\end{code}

That is, a stream is again a pair of a hidden state, |'s| (only
known dynamically, i.e., |'s code|),
and a step function, but the step function does not return
|stream_shape| values (of dynamic |'a|s and |'s|s) but accepts an
extra argument (the continuation) to pass such values to. The step
function returns whatever (generic type |'w|, only known dynamically) the continuation
returns.

The variants of the |stream_shape| are now known when |step|
calls its continuation, which happens at code-generation time. The
|map| combinator becomes
\begin{code}
let map : ('a code -> 'b code) ->
          'a st_stream -> 'b st_stream =
  fun f (s,step) ->
    let new_step s k = step s @@ function
    | Nil        -> k Nil
    | Cons (a,t) -> .<let a' = .~(f a) in
                     .~(k @@ Cons (.<a'>., t))>.
    in (s,new_step)
\end{code}
taking into account that |step|, instead of returning the result, calls a
continuation on it. Although the data-type |stream_shape| remains, its
construction and  pattern-matching now happen at code-generation
time, i.e., statically. As another example, the |fold| combinator becomes:
\begin{code}
let fold : ('z code -> 'a code -> 'z code) ->
            'z code -> 'a st_stream -> 'z code
 = fun f z (s,step) ->
 .<let rec loop z s = .~(step .<s>. @@ function
   | Nil        -> .<z>.
   | Cons (a,t) -> .<loop .~(f .<z>. a) .~t>.)
   in loop .~z .~s>.
\end{code}
Our running example pipeline, summing the squares of all elements of
a sample array, now generates the following code
\begin{code}
val c : int code = .<
  let rec loop_1 z_2 s_3 =
    match s_3 with
    | (i_4,arr_5) ->
        if i_4 < (Array.length arr_5)
        then
          let el_6 = arr_5.(i_4) in
          let a'_7 = el_6 * el_6 in
          loop_1 (z_2 + a'_7) ((i_4 + 1), arr_5)
        else z_2 in
  loop_1 0 (0, [|0;1;2;3;4|])>.
\end{code}
In stark contrast with the naive staging
of \S\ref{sec:simple-staging}, the generated code has no traces of
the |stream_shape| data type.
Although the data type is still constructed and deconstructed,
the corresponding overhead is shifted from the generated code to the
code-generator. Generating code may take a bit longer but the
result is more efficient. For full fusion, we will need to shift
overhead to the generator two more times.

\subsection{Fusing the Stream State}
\label{sec:state-fusing}

Although we have removed the most noticeable repeated
construction and deconstruction of the |stream_shape| data type,
the abstraction overhead still remains. The main loop in the generated code
pattern-matches on the current state, which is the pair of the index
and the array. The recursive invocation of the loop packs the index
and the array back into a pair. Our task is to deforest the pair
away. This seems rather difficult, however: the state is being updated
on every iteration of the loop, and the loop structure (e.g., number
of iterations) is generally not statically known.  Although it is the
(statically known) |step| function that computes the updated state,
the state has to be threaded through the fold's |loop|, which
treats it as a black-box piece of code. The
fact it is a pair cannot be exploited and, hence, the overhead cannot
be shifted to the generator. There is a way out, however.  It requires
a non-trivial step:
The threading of the state through the loop can be eliminated if the
state is mutable.

The step function no longer has to return (strictly speaking: pass to its
continuation) the updated state: the
update happens in place. Therefore, the state no longer
has to be annotated as dynamic---its structure can be known to the
generator. Finally, in order to have the appropriate operator
allocate the reference cell for the array index, we need to employ
the let-insertion technique \cite{bondorf-improving}, by also
using continuation-passing style for
the initial state. The definition of the stream type (|'a st_stream|)
now becomes:
\begin{code}
type 'a st_stream =
 exists's.
   (forall'w. ('s -> 'w code) -> 'w code) *
   (forall'w. 's ->
     (('a code,unit) stream_shape -> 'w code) ->
         'w code)
\end{code}
That is, a stream is a pair of an |init| function and a |step| function.
The |init| function implicitly hides a state: it knows how to call a continuation
(that accepts
a static state and returns a generic dynamic value, |'w|) and returns
the result of the continuation. The |step| function is much like
before, but operating on a statically-known state (or more correctly, a
hidden state with a statically-known structure).

The new |of_arr| combinator demonstrates
the let-insertion (the allocation of the reference cell for the
current array index) in |init|, and the in-place update of the state (the |incr|
operation):
\begin{code}
let of_arr : 'a array code -> 'a st_stream =
  let init arr k =
    .<let i = ref 0 and
          arr = .~arr in .~(k (.<i>.,.<arr>.))>.
  and step (i,arr) k =
   .<if !(.~i) < Array.length .~arr
      then
       let el = (.~arr).(!(.~i)) in
       incr .~i;
       .~(k @@ Cons (.<el>., ()))
      else .~(k Nil)>.
  in
  fun arr -> (init arr,step)
\end{code}
Once again, until now the state of the |of_arr| stream had the type
|(int * 'a array) code|. It has become |int ref code * 'a array code|,
the statically known pair of two code values. The construction and
deconstruction of that pair now happens at code-generation time.

The earlier |map| combinator did not even look at the current state (nor
could it), therefore its code remains unaffected by the change in the
state representation. The |fold| combinator no longer has to thread
the state through its loop:
\begin{code}
let fold : ('z code -> 'a code -> 'z code) ->
            'z code -> 'a st_stream -> 'z code
 = fun f z (init,step) ->
 init @@ fun s ->
 .<let rec loop z = .~(step s @@ function
   | Nil        -> .<z>.
   | Cons (a,_) -> .<loop .~(f .<z>. a)>.)
   in loop .~z>.
\end{code}
It obtains the
state from the initializer and passes it to the step function, which
knows its structure. The generated code for the running-example
stream-processing pipeline is:
\begin{code}
val c : int code = .<
  let i_8 = ref 0
  and arr_9 = [|0;1;2;3;4|] in
  let rec loop_10 z_11 =
    if ! i_8 < Array.length arr_9
    then
      let el_12 = arr_9.(! i_8) in
      incr i_8;
      let a'_13 = el_12 * el_12 in
      loop_10 (z_11+a'_13)
    else z_11 in
  loop_10 0>.
\end{code}
The resulting code shows the absence of any overhead. All intermediate
data structures have been eliminated. The code is what we could expect
to get from a competent OCaml programmer.

\subsection{Generating Imperative Loops}
\label{sec:imperative-loops}

It seems we have achieved our goal. The library (extended for
filtering, zipping, and nested streams) can be used in (Meta)OCaml
practice. It relies, however, on tail-recursive function calls. These
may be a good fit for OCaml,\footnote{Actually, our benchmarking reveals
  that for- and while-loops are currently faster even in OCaml.}
but not for Java or Scala. (In Scala, tail-recursion is only
supported with significant run-time overhead.) The fastest way to
iterate is to use the native while-loops, especially in Java or
Scala. Also, the dummy |('a code,unit) stream_shape| in
the |'a st_stream| type looks
odd: the |stream_shape| data type has become artificial.
Although |unit| has no effect on generated code, it is
less than pleasing aesthetically to need a placeholder type in our
signature. For these reasons, we embark on one last transformation.

The last step of stream staging is driven by several insights. First
of all, most languages provide two sorts of imperative loops: a
general while-loop and the more specific, and often more efficient (at
least in OCaml) for-loops. We would like to be able to generate
for-loops if possible, for instance, in our running example.
However, with added subranging or zipping (described in detail
in \S\ref{sec:implementation}, below) the pipeline can no longer be represented as an
OCaml for-loop, which cannot accommodate extra termination
tests. Therefore, the stream producer should not commit to any particular
loop representation. Rather, it has to collect all the needed
information for loop generation, but leave the actual generation to the
stream consumer, when the entire pipeline is known. Thus the
stream representation type becomes as follows:

\begin{code}
type ('a,'s) producer_t =
  | For    of
     {upb:   's -> int code;
      index: 's -> int code -> ('a -> unit code) ->
                   unit code}
  | Unfold of
     {term: 's -> bool code;
      step: 's -> ('a -> unit code) -> unit code}
 and 'a st_stream =
   exists's. (forall'w. ('s -> 'w code) -> 'w code) *
              ('a,'s) producer_t
 and 'a stream = 'a code st_stream
\end{code}

That is, a stream type is a pair of an |init| function (which, as before,
has the ability to call a continuation with a hidden state) and an encoding
of a producer. We distinguish two sorts of producers: a producer that can be driven
by a for-loop or a general ``unfold'' producer. Each of them supports two
functions. A for-loop producer carries the exact upper bound, |upb|, for
the loop index variable and the |index| function that returns the stream element
given an index. For a general producer,
we refactor (with an eye for the while-loop) the earlier representation
\begin{code}
(('a code,unit) stream_shape -> 'w code) -> 'w code
\end{code}
into two components: the termination test, |term|, producing
a dynamic |bool| value (if the test yields
|false| for the current state, the loop is finished) and the |step| function,
to produce a new stream element and advance the state. We also used
another insight: the imperative-loop--style of the processing pipeline
makes it unnecessary (moreover, difficult) to be passing around the
consumer (|fold|) state from one iteration to another. It is easier to
accumulate the state in a mutable cell. Therefore, the answer type of
the |step| and |index| functions can be |unit code| rather than |'w code|.

There is one more difference from the earlier staged stream, which is a bit
harder to see. Previously, the stream value was annotated as
dynamic: we really cannot know before running the
pipeline what the current element is. Now, the value produced by the |step| or
|index| functions has the type |'a| without any |code| annotations,
meaning that it is statically known! Although the value of the current
stream element is determined only when the pipeline is run, its
structure can be known earlier. For example, the new type lets the
producer yield a pair of values: even though the values themselves are
annotated as dynamic (of a |code| type) the fact that it is a pair can be
known statically. We use this extra flexibility of the more general
stream value type extensively in \S\ref{sec:take}.

We can now see the new design in action.
The stream producer |of_arr| is surely the for-loop-style producer:
\begin{code}
let of_arr : 'a array code -> 'a stream = fun arr ->
  let init k = .<let arr = .~arr in .~(k .<arr>.)>.
  and upb arr = .<Array.length .~arr - 1>.
  and index arr i k =
      .<let el = (.~arr).(.~i) in .~(k .<el>.)>.
  in (init, For {upb;index})
\end{code}
In contrast, the unfold combinator
\begin{code}
let unfold : ('z code -> ('a * 'z) option code) ->
             'z code -> 'a stream = ...
\end{code}
is an |Unfold| producer.

Importantly, a producer that starts as a for-loop may later be converted to a
more general while-loop producer, (so as to tack on extra
termination tests---see |take| in \S\ref{sec:take}). Therefore, we need the
conversion function
\begin{code}
let for_unfold : 'a st_stream -> 'a st_stream=  function
 | (init,For {upb;index}) ->
    let init k = init @@ fun s0 ->
            .<let i = ref 0 in .~(k (.<i>.,s0))>.
    and term (i,s0)   = .<!(.~i) <= .~(upb s0)>.
    and step (i,s0) k =
      index s0 .<!(.~i)>. @@
            fun a -> .<(incr .~i; .~(k a))>.
    in (init, Unfold {term;step})
 | x -> x
\end{code}
used internally within the library.

The stream mapping operation composes the mapping function with the
|index| or |step|: transforming, as before, the produced value
``in-flight'', so to speak.
\begin{code}
let rec map_raw: ('a -> ('b -> unit code) -> unit code)
                  -> 'a st_stream -> 'b st_stream =
 fun tr -> function
 | (init,For ({index;_} as g)) ->
     let index s i k = index s i @@ fun e -> tr e k in
     (init, For {g with index})
 | (init,Unfold ({step;_} as g)) ->
     let step s k = step s @@ fun e -> tr e k in
     (init, Unfold {g with step})
\end{code}
We have defined |map_raw| with the general type (to be used later,
e.g., in \S\ref{sec:take}); the familiar |map| is a special case:
\begin{code}
let map : ('a code -> 'b code) -> 'a stream -> 'b stream
  = fun f str -> map_raw (fun a k ->
      .<let t = .~(f a) in .~(k .<t>.)>.) str
\end{code}
The mapper |tr| in
|map_raw| is in the continuation-passing style with the
|unit code| answer-type. This allows us to perform let-insertion
\cite{bondorf-improving}, binding the mapped value to a variable, and
hence avoiding the potential duplication of the mapping operation.

As behooves pull-style streams, the consumer at the end of the
pipeline generates the loop to drive
the iteration. Yet we do manage to generate for-loops, characteristic of
push-streams, see \S\ref{sec:fusion-problem}.
\begin{code}
let rec fold_raw :
 ('a -> unit code) -> 'a st_stream -> unit code
 = fun consumer -> function
   | (init,For {upb;index}) ->
         init @@ fun sp ->
           .<for i = 0 to .~(upb sp) do
              .~(index sp .<i>. @@ consumer)
             done>.
   | (init,Unfold {term;step}) ->
         init @@ fun sp ->
           .<while .~(term sp) do
             .~(step sp @@ consumer)
           done>.
\end{code}
It is simpler (especially when we add nesting later) to implement a
more general |fold_raw|, which feeds the eventually produced
stream element to the given imperative |consumer|.
The ordinary |fold| is a wrapper that provides such a consumer,
accumulating the result in
a mutable cell and extracting it at the end.
\begin{code}
let fold : ('z code -> 'a code -> 'z code) ->
           'z code -> 'a stream -> 'z code
= fun f z str ->
  .<let s = ref .~z in
    (.~(fold_raw
         (fun a -> .<s := .~(f .<!s>. a)>.)
         str);
    !s)>.
\end{code}

The generated code for our running example is:
\begin{code}
val c : int code = .<
  let s_1 = ref 0 in
  let arr_2 = [|0;1;2;3;4|] in
   for i_3 = 0 to (Array.length arr_2) - 1 do
     let el_4 = arr_2.(i_3) in
     let t_5 = el_4 * el_4 in s_1 := !s_1 + t_5
   done;
  ! s_1>.
\end{code}
This code could not be better. It is what we expect an OCaml programmer to
write, and, furthermore, such code performs ultimately
well in Scala, Java and other languages.
We have achieved our goal---for simple pipelines, at least.

\section{Full Library}
\label{sec:implementation}

The previous section presented our approach of eliminating all
abstraction overhead of a stream library through the creative use of
staging---generating code of hand-written quality and efficiency.
However, a full stream library has more combinators than we have
dealt with so far. This section describes the remaining facilities:
filtering, sub-ranging, nested streams and parallel streams
(zipping).
Consistently achieving deforestation and high performance in the
presence of all these features is a challenge.
We identify three concepts of stream processing that drive our effort:
the rate of production and consumption of stream elements
(\emph{linearity} and filtering---\S\ref{sec:filter-nest}),
size-limiting a stream (\S\ref{sec:take}), and processing multiple
streams in tandem (zipping---\S\ref{sec:zip}). We conclude our core
discussion with a theorem of eliminating all overhead.


\subsection{Filtered and Nested Streams}
\label{sec:filter-nest}

Our library is primarily based on the design presented at the end of
\S\ref{sec:overhead-elim}. Filtering and nested streams (|flat_map|)
require an extension, however, which lets us treat filtering
and flat-mapping uniformly.

Let us look back at this design.  It centers on two operations, |term|
and |step|: forgetting for a moment the staging annotations, |term s|
decides whether the stream still continues, while |step s| produces
the current element and advances the state. Exactly one stream element
is produced per advance in state. We call such streams
\emph{linear}. They have many useful algebraic properties, especially when it
comes to zipping. We will exploit them in \S\ref{sec:zip}.

Clearly the |of_arr| stream producer and the more general |unfold|
producers build linear streams. The |map| operation preserves the
linearity. What destroys it is filtering and nesting. In the filtered
stream \lstinline{prod |> filter p}, the advancement of the |prod|
state is no longer always accompanied by the production of the stream
element: if the filter predicate |p| rejects the element, the pipeline
will yield nothing for that iteration. Likewise, in the nested stream
\lstinline{prod |> flat_map (fun x -> inner_prod x)}, the advancement
of the |prod| state may lead to zero, one, or many stream elements
given to the pipeline consumer.

Given the importance of linearity (to be seen in full in
\S\ref{sec:zip}) we keep track of it in the stream representation. We
represent a non-linear stream as a composition of an always-linear
producer with a non-linear transformer:
\begin{code}
type card_t = AtMost1 | Many

type ('a,'s) producer_t =
  | For    of
     {upb:   's -> int code;
      index: 's -> int code -> ('a -> unit code) ->
                   unit code}
  | Unfold of
     {term: 's -> bool code;
      card: card_t;
      step: 's -> ('a -> unit code) -> unit code}
 and 'a producer =
   exists's. (forall'w. ('s -> 'w code) -> 'w code) *
              ('a,'s) producer_t
 and 'a st_stream =
  | Linear of 'a producer
  | Nested of exists'b. 'b producer * ('b -> 'a st_stream)
 and 'a stream = 'a code st_stream
\end{code}
The difference from the earlier representation in
\S\ref{sec:overhead-elim} is the addition of a
sum data type with variants |Linear| and |Nested|,
for linear and nested streams. We also
added a cardinality marker to the general producer, noting if it
generates possibly many elements or at most one.

The |flat_map| combinator adds a non-linear transformer to the
stream (recursively descending into the already nested stream):
\begin{code}
let rec flat_map_raw :
  ('a -> 'b st_stream) -> 'a st_stream -> 'b st_stream =
 fun tr -> function
 | Linear prod         -> Nested (prod,tr)
 | Nested (prod,nestf) ->
    Nested (prod,fun a -> flat_map_raw tr @@ nestf a)

let flat_map :
  ('a code -> 'b stream) -> 'a stream -> 'b stream =
  flat_map_raw
\end{code}
The |filter| combinator becomes just a particular case of flat-mapping: nesting of
a stream that produces at most one element:
\begin{code}
let filter : ('a code -> bool code) ->
             'a stream -> 'a stream = fun f ->
  let filter_stream a =
  ((fun k -> k a),
   Unfold {card = AtMost1; term = f;
           step = fun a k -> k a})
  in flat_map_raw (fun x -> Linear (filter_stream x))
\end{code}
The addition of recursively |Nested| streams requires an adjustment
of the earlier, \S\ref{sec:overhead-elim},
|map_raw| and |fold| definitions to recursively descend down the nesting. The
adjustment is straightforward; please see the accompanying source code
for details. The adjusted |fold| will generate nested loops for
nested streams.

\subsection{Sub-Ranging and Infinite Streams}
\label{sec:take}

The stream combinator |take| limits the size of the stream:
\begin{code}
val take : int code -> 'a stream -> 'a stream
\end{code}
For example, |take .<10>. str| is a stream of the first 10 elements of
|str|, if there are that many.  It is the |take| combinator that lets us
handle conceptually infinite streams. Such infinite streams are easily
created with |unfold|: for example, |iota n|, the stream of all
natural numbers from |n| up:
\begin{code}
let iota n = unfold (fun n -> .<Some (.~n,.~n+1)>.) n
\end{code}

The implementation of |take| demonstrates and justifies design
decisions that might have seemed arbitrary earlier. For example,
distinguishing linear streams and indexed, for-loop--style producers
in the representation type pays off. In a linear stream pipeline, the
number of elements at the end of the pipeline is the same as the
number of produced elements. Therefore, for a linear stream, |take|
can impose the limit close to the production. The for-loop-style
producer is particularly easy to limit in size: we merely need to
adjust the upper bound:
\begin{code}
let take = fun n -> function
 | Linear (init, For {upb;index}) ->
    let upb s = .<min (.~n-1) .~(upb s)>. in
    Linear (init, For {upb;index})
 ...
\end{code}
Limiting the size of a non-linear stream is slightly more
complicated:
\begin{code}
let take = fun n -> function
 ...
 | Nested (p,nestf) ->
    Nested (add_nr n (for_unfold p),
     fun (nr,a) ->
      map_raw (fun a k -> .<(decr .~nr; .~(k a))>.) @@
      more_termination .<! .~nr > 0>. (nestf a))
\end{code}
The idea is straightforward: allocate a reference cell |nr|
with the remaining element count (initially |n|), add the check
|!nr > 0| to the termination condition of the stream producer, and
arrange to decrement the |nr| count at the end of the
stream. Recall, for a non-linear stream---a composition of several
producers---the count of eventually produced elements may differ
arbitrarily from the count of the elements emitted by the first
producer. A moment of thought shows that the range check |!nr > 0|
has to be added not only to the first producer but to the producers of
all nested substreams: this is the role of
function |more_termination| (see the accompanying code for its
definition) in the fragment above. The operation
|add_nr| allocates cell |nr| and adds the termination condition to
the first producer. Recall that, since for-loops in OCaml cannot take
extra termination conditions, a for-loop-style producer has to be first
converted to a general unfold-style producer, using |for_unfold|, which
we defined in \S\ref{sec:overhead-elim}. The operation |add_nr|
(definition not shown) also adds |nr| to the produced value: The result of
|add_nr n (for_unfold p)| is of type
|(int ref code,'a code) st_stream|. Adding the operation to
decrement |nr| is conveniently done with |map_raw|
from \S\ref{sec:overhead-elim}.
We, thus, now see the use for the more general (|'a| and not just |'a code|)
stream type and the general stream mapping function.

\subsection{zip: Fusing Parallel Streams}
\label{sec:zip}

This section describes the most complex operation: handling two
streams in tandem, i.e., zipping:
\begin{code}
val zip_with   : ('a code -> 'b code -> 'c code) ->
                 ('a stream -> 'b stream -> 'c stream)
\end{code}
Many stream libraries lack this operation: first, because zipping is
practically impossible with push streams, due to inherent complexity,
as we shall see shortly. Linear streams and the general |map_raw|
operation turn out to be important abstractions that make the problem
tractable.

One cause of the complexity of |zip_with| is the need to consider many
special cases, so as to generate code of hand-written
quality. All cases share the operation of combining the elements
of two streams to obtain the element of the zipped stream. It is
convenient to factor out this operation:
\begin{code}
val zip_raw: 'a st_stream -> 'b st_stream ->
             ('a * 'b) st_stream

let zip_with f str1 str2 =
   map_raw (fun (x,y) k -> k (f x y)) @@
   zip_raw str1 str2
\end{code}
The auxiliary |zip_raw| builds a stream of pairs---statically known
pairs of dynamic values. Therefore, the overhead of constructing and
deconstructing the pairs is incurred only once, in the generator. There is
no tupling in the generated code.

The |zip_raw| function is a dispatcher for various special cases, to
be explained below.
\begin{code}
let rec zip_raw str1 str2 = match (str1,str2) with
  | (Linear prod1, Linear prod2) ->
      Linear (zip_producer prod1 prod2)
  | (Linear prod1, Nested (prod2,nestf2)) ->
      push_linear (for_unfold prod1)
                  (for_unfold prod2,nestf2)
  | (Nested (prod1,nestf1), Linear prod2) ->
      map_raw (fun (y,x) k -> k (x,y)) @@
      push_linear (for_unfold prod2)
                  (for_unfold prod1,nestf1)
  | (str1,str2) ->
      zip_raw (Linear (make_linear str1)) str2
\end{code}

The simplest case is zipping two linear streams. Recall, a linear
stream produces exactly one element when advancing the state. Zipped
linear streams, hence, yield a linear stream that produces a pair of
elements by advancing the state of both argument streams exactly once.
The pairing of the stream advancement is especially efficient for
for-loop--style streams, which share a common state, the index:
\begin{code}
let rec zip_producer:
  'a producer -> 'b producer -> ('a * 'b) producer =
 fun p1 p2 -> match (p1,p2) with
 | (i1,For f1), (i2,For f2) ->
     let init k =
       i1.init @@ fun s1 ->
       i2.init @@ fun s2 -> k (s1,s2)
     and upb (s1,s2) = .<min .~(f1.upb s1)
                             .~(f2.upb s2)>.)
     and index fun (s1,s2) i k =
       f1.index s1 i @@ fun e1 ->
       f2.index s2 i @@ fun e2 -> k (e1,e2)
     in (init, For {upb;index})
  | (* elided *)
\end{code}

In the general case, |zip_raw str1 str2| has to determine how to advance
the state of |str1| and |str2| to produce one element of the zipped
stream: the pair of the current elements of |str1| and
|str2|. Informally, we have to reason all the way from the production of an
element to the advancement of the state. For linear streams, the
relation between the current element and the state is one-to-one. In
general, the state of the two components of the zipped stream advance
at different paces.  Consider the following sample streams:
\begin{code}
let stre = of_arr arr1
           |> filter (fun x -> .<.~x mod 2 = 0>.)
let strq = of_arr arr2
           |> map (fun x -> .<.~x * .~x>.)
let str2 = of_arr arr1
           |> flat_map (fun _ -> of_arr .<[|1;2]>.)
let str3 = of_arr arr1
           |> flat_map (fun _ -> of_arr .<[|1;2;3]>.)
\end{code}
To produce one element of |zip_raw stre strq|, the state of |stre|
has to be advanced a statically-unknown number of times.
Zipping nested streams is even harder---e.g.,
|zip_raw str2 str3|, where the states advance in complex
patterns and the end of the inner stream of |str2| does not align with
the end of the inner stream in |str3|.

Zipping simplifies if one of the streams is linear, as in
|zip_raw stre| |strq|. The key insight is
to advance the linear stream |strq| after we are sure to have obtained the
element of the non-linear stream |stre|. This idea is elegantly
realized as mapping of the step function of |strq| over |stre|
(the latter, is, recall, |int stream|, which is |int code st_stream|),
obtaining the desired zipped |(int code, int code)| |st_stream|:
\begin{code}
map_raw (fun e1 k ->
         strq.step sq (fun e2 -> k (e1,e2))) stre
\end{code}

The above code is an outline: we have to initialize |strq| to obtain its
state |sq|, and we need to push the termination condition of |strq|
into |stre|. Function |push_linear| in the accompanying code
takes care of all these details.

The last and most complex case is zipping two non-linear streams. Our solution is
to convert one of them to a linear stream, and then use the approach just
described. Turning a non-linear stream to a producer
involves ``reifying'' a stream: converting an |'a stream| data type to
essentially a |(unit -> 'a option) code| function, which, when called,
reports the new element or the end of the stream. We have to create a
closure and generate and deconstruct the intermediate data
type |'a option|. There is no way around this: in one form or another,
we have to capture the non-linear stream's continuation. The human
programmer will have to do the same---this is precisely what makes zipping so difficult
in practice. Our library reifies only one of the two zipped
streams, without relying on tail-call optimization, for
maximum portability.

\subsection{Elimination of All Overhead, Formally}

Sections \ref{sec:overview}, above, and \ref{sec:experiments}, below,
demonstrate the elimination of abstraction overhead on selected
examples and benchmarks. We now state how and why the overhead is
eliminated in all cases.

We call the higher-order arguments of |map|, |filter|, |zip_with|,
etc. ``user-generators'': they are specified by the library user and
provide per-element stream processing.

\begin{theorem}\label{thm:expr}
Any well-typed pipeline generator---built by composing a stream
producer, Fig.\ref{f:interface}, with an
arbitrary combination of transformers followed by a reducer---terminates,
provided the user-generators do. The resulting code---with the sole
exception of pipelines zipping two flat-mapped
streams---constructs no data structures beyond those
constructed by the user-generators.
\end{theorem}

Therefore, if the user generators proceed without
construction/allocation, the entire pipeline, after the initial
set-up, runs without allocations. The only exception is the zipping of
two streams that are both made by flattening inner streams. In
this case, the rate-adjusting allocation is inevitable, even in
hand-written code, and is not considered overhead.

\emph{Proof sketch:} The proof is simple, thanks to the explicitness
of staging and treating the generated code as an opaque value that
cannot be deconstructed and examined. Therefore, the only tuple
construction operations in the generated code are those that we have
explicitly generated. Hence, to prove our theorem, we only have to
inspect the brackets that appear in our library implementation,
checking for tuples or other objects.

\section{Experiments}
\label{sec:experiments}

We evaluated our approach on several benchmarks from past literature,
measuring the iteration throughput:
\begin{bullets}
 \item \textbf{sum}: the simplest
\lstinline{of_arr arr |> sum} pipeline, summing the elements of an array;
 \item \textbf{sumOfSquares}: our running example from
   \S\ref{sec:simple-staging} on;
 \item \textbf{sumOfSquaresEven}: the sumOfSquares benchmark with added
   filter, summing the squares of only the even array elements;
 \item \textbf{cart}: $\sum x_iy_j$, using |flat_map| to build the outer-product stream;
 \item \textbf{maps}: consecutive map operations with integer
   multiplication;
 \item \textbf{filters}: consecutive filter operations using
   integer comparison;
 \item \textbf{dotProduct}: compute dot product of two arrays using
|zip_with|;
 \item \textbf{flatMap\_after\_zipWith}: compute $\sum (x_i+x_i)y_j$,
   like cart above, doubling the |x| array via |zip_with (+)| with itself;
 \item \textbf{zipWith\_after\_flatMap}: |zip_with| of two streams one
   of which is the result of |flat_map|;
 \item \textbf{flat\_map\_take}: |flat_map| followed by |take|.
\end{bullets}
The source code of all benchmarks is available at the project's
repository and the OCaml versions are also listed in
Appendix~D of the extended version.
Our
benchmarks come from the sets by Murray et al.~\cite{murray_steno:_2011} and
Coutts et al.~\cite{coutts_stream_2007}, to which we added more complex
combinations (the last three on the list above). (The Murray and Coutts sets
also contain a few more simple operator combinations, which we omit for conciseness,
as they share the
performance characteristics of other benchmarks.)

The staged code was generated using our library (\strymonas{}), with MetaOCaml on the
OCaml platform and LMS on Scala, as detailed below.
As one basis of
comparison, we have implemented all benchmarks using
the streams libraries available on each platform\footnote{\relax
We restrict our attention to
the closest feature-rich apples-to-apples comparables: the industry-standard
libraries for OCaml+JVM languages. We also report qualitative comparisons
in \S\ref{sec:related}.}:
Batteries~\footnote{Batteries is the widely used ``extended
  standard'' library in
  OCaml~\url{http://batteries.forge.ocamlcore.org/}.}
in OCaml and the standard Java 8 and Scala streams. As there is not a unifying
module that implements all the combinators we employ, we use data type
conversions where possible. Java 8 does not support a |zip| operator,
hence some benchmarks are missing for that setup.\footnote{One could
emulate \texttt{zip} using
\texttt{iterator} from Java 8 push-streams---at significant
drop in performance. This encoding also markedly differs from
the structure of our other stream implementations.}

As the baseline and the other basis of comparison, we have hand-coded
all the benchmarks, using high-performance, imperative code, with
|while| or index-based |for|-loops, as applicable.
In \scala{} we use only |while|-loops as they are the analogue of
imperative iterations; |for|-loops in Scala operate over |Range|s
and have worse performance. In fact, in one case we had
to re-code the hand-optimized loop upon discovering that it was
not as optimal as we thought: the library-generated code
significantly outperformed it!

\paragraph*{Input:} All tests were run with the same input set. For the
\textbf{sum}, \textbf{sumOfSquares}, \textbf{sumOfSquaresEven}, \textbf{maps},
\textbf{filters} we used an array of $N = 100,000,000$ small integers:
$x_i = i\ \mathrm{mod}\ 10$. The
\textbf{cart} test iterates over two arrays. An outer one of $10,000,000$
integers and an inner one of $10$. For the \textbf{dotProduct} we used
$10,000,000$ integers, for the \textbf{flatMap\_after\_zipWith} $10,000$, for
the \textbf{zipWith\_after\_flatMap} $10,000,000$ and for the
\textbf{flat\_map\_take} $N$ numbers sub-sized by $20\%$ of $N$.

\paragraph*{Setup:} The system we use runs an x64 OSX El Capitan 10.11.4
operating system on bare metal. It is equipped with a 2.7 GHz Intel Core i5
CPU (I5-5257U) having 2 physical and 2 logical cores. The total memory of the
system is 8 GB of type 1867 MHz DDR3. We use version build 1.8.0\_65-b17 of the
Open JDK. The compiler versions of our setup are presented in the table below:
\begin{center}
\begin{tabular}{ c c c }
  \toprule
  Language & Compiler & Staging \\
  \midrule
  Java & Java 8 (1.8.0\_65)  & \textemdash \\
  Scala & 2.11.2 & LMS 0.9.0 \\
  OCaml & 4.02.1 & BER MetaOCaml N102 \\
  \bottomrule
\end{tabular}
\end{center}

\addtolength{\abovecaptionskip}{-8mm}
\addtolength{\belowcaptionskip}{-1.5mm}

\begin{figure*}
\centering
\begin{subfigure}{0.9\textwidth}
  \centering
  \includegraphics[width=.98\linewidth]{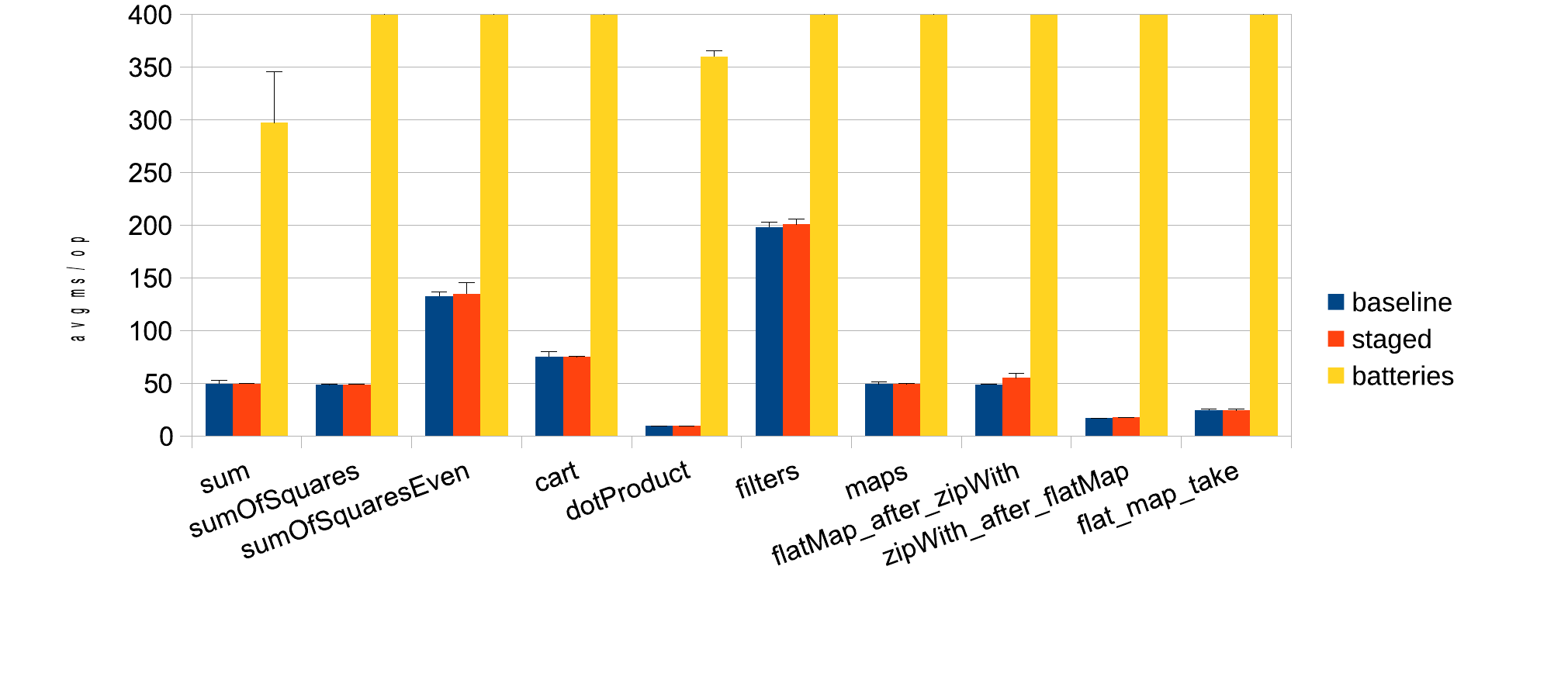}
\end{subfigure}
\caption{OCaml microbenchmarks in msec / iteration (avg. of 30, with mean-error bars shown). ``Staged'' is our library (\strymonas{}). The figure is truncated: OCaml batteries take more than 60sec (per iteration!) for some complex benchmarks.}
\label{fig:microbenchmarks1}
\end{figure*}

\begin{figure*}
\centering
\begin{subfigure}{0.9\textwidth}
  \centering
  \includegraphics[width=.98\linewidth]{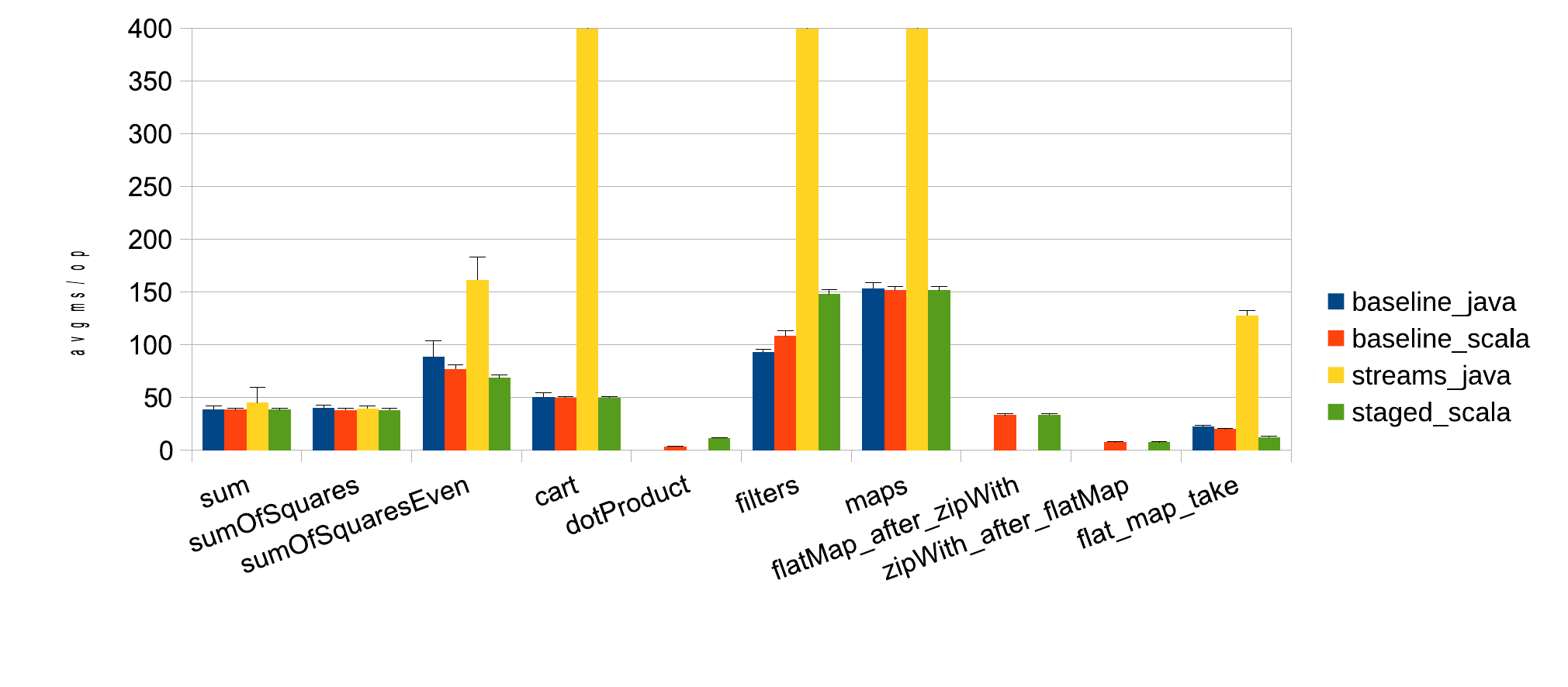}
\end{subfigure}
\caption{JVM microbenchmarks (both Java and Scala) in msec / iteration (avg. of 30, with mean-error bars shown). ``Staged\_scala'' is our library (\strymonas{}). The figure is truncated.}
\label{fig:microbenchmarks2}
\end{figure*}

\addtolength{\abovecaptionskip}{+8mm}
\addtolength{\belowcaptionskip}{+1mm}

\paragraph*{Automation:} For Java and \scala{} benchmarks we used the Java
Microbenchmark Harness (JMH)~\cite{aleksey_shipilev_openjdk} tool: a
benchmarking tool for JVM-based languages that is part of the OpenJDK. JMH is an
annotation-based tool and takes care of all intrinsic details of the execution
process. Its goal is to produce as objective results as possible. The JVM
performs JIT compilation (we use the C2 JIT compiler) so the benchmark author
must measure execution time after a certain warm-up period to wait for transient
responses to settle down. JMH offers an easy API to achieve that. In our
benchmarks we employed 30 warm-up iterations and 30 proper iterations. We also
force garbage collection before benchmark execution and between runs. All OCaml code
was compiled with |ocamlopt| into machine code. In particular, the
MetaOCaml-generated code was saved into a file, compiled, and then
benchmarked in isolation. The test harness invokes the compiled
executable via |Sys.command|, which is
not included in the results. The harness calculates the average
execution time, computing the mean error and standard deviation using
the Student-T distribution. The same method is employed
in JMH.
For all tests, we do not measure the time needed to initialize
data-structures (filling arrays), nor the run-time compilation cost of
staging. These costs are constant (i.e., they become proportionally
insignificant for larger inputs or more iterations) and they were
small, between 5 and 10ms, for all our runs.

\paragraph*{Results:} In Figures~\ref{fig:microbenchmarks1} and
~\ref{fig:microbenchmarks2} we present the results of our experiments divided
into two categories: a) the OCaml microbenchmarks of baseline, staged and
batteries experiments and b) the JVM microbenchmarks. The JVM diagram
contains the baselines for both Java and Scala. Shorter bars are better.
Recall that all ``baseline'' implementations are carefully hand-optimized code.

As can be seen, our staged library achieves extremely high
performance, matching hand-written code (in either OCaml, Java, or
Scala) and outperforming other library options by orders of
magnitude. Notably, the highly-optimized Java 8 streams are more than
10x slower for perfectly realistic benchmarks, when those do not
conform to the optimal pattern (linear loop) of push streams.

\section{Related Work}
\label{sec:related}

The literature on stream library designs is rich. Our approach is the first
to offer full generality while eliminating processing overhead. We discuss
individual related work in more detail next.

One of the earliest stream libraries that rely on staging is Common Lisp's
SERIES \cite{lisp-series,waters_series_1991}, which extensively relies on Lisp
macros to interpret a subset of Lisp code as a stream EDSL. It builds a data
flow graph and then compiles it into a single loop. It can handle filtering,
multiple producers and consumers, but not nested streams. The (over)reliance on
macros may lead to surprises since the programmer might not be aware that what
looks like CL code is actually a DSL, with a slightly different semantics and
syntax. An experimental Pipes package \cite{lisp-pipes} attempts to re-implement
and extend SERIES, using, this time, a proper EDSL. Pipes extends SERIES by
allowing nesting, but restricts zipping to simple cases. It was posited
that ``arbitrary outputs per input, multiple consumers, multiple producers:
choose two'' \cite{lisp-pipes}. Pipes ``almost manages'' (according to its
author) to implement all three features. Our library demonstrates the
conjecture is false by supporting all three facilities in full generality
and with high performance.

Lippmeier et al.~\cite{Lippmeier_dataflow_2013} present a line of work
based on SERIES. They aim to transform first-order, non-recursive, synchronous,
finite data-flow programs into fused pipelines. They derive inspiration from
traditional data-flow languages like Lustre~\cite{halbwachs_synchronous_1991}
and Lucid Synchrone~\cite{pouzet_lucid_2006}. In contrast, our library
supports a greater range of fusible combinators, but for bulk data processing.





Haskell has lazy lists, which seem to offer incremental processing
by design. Lazy lists cannot express pipelines that require
side-effects such as reading or writing files.\footnote{We disregard
  the lazy IO misfeature
  \cite{iteratees}.} The all-too-common memory leaks point out that
lazy lists do not offer, again by design, stream fusion.
Overcoming the drawbacks of lazy lists, coroutine-like iteratees
\cite{iteratees} and many of their reimplementations support incremental
processing even in the presence of effects, for nested
streams and for several consumers and producers. Although iteratees avoid
intermediate streams they still suffer large overheads for captured
continuations, closures, and coroutine calls.


Coutts et al.~\cite{coutts_stream_2007} proposed \emph{Stream
Fusion} (the approach that has become associated with this fixed term), building on previous work (|build|/|foldr|~\cite{gill_shortcut_1993}
and |destroy|/|unfoldr|~\cite{svenningsson_shortuct_2002}) by fusing maps,
filters, folds, zips and nested lists. The approach relies on
the rewrite GHC \textsc{Rules}. Its notable contribution is the support for
stream filtering. In that approach there is no specific treatment of
linearity. The Coutts et al. stream fusion supports zipping, but only in simple cases (no zipping
of nested, subranged streams). Finally, the Coutts et al. approach does not fully fuse
pipelines that contain nested streams (|concatMap|). The reason is that the
stream created by the transformation of |concatMap| uses an internal function
that cannot by optimized by GHC by employing simple case reduction. The problem
is presented very concisely by Farmer et al. in the
\emph{Hermit in the Stream} work~\cite{farmer_hermit_2012}.

The application of HERMIT~\cite{farmer_hermit_2012} to streams
\cite{farmer_hermit_2014} fixes the shortcomings of the Coutts et al. Stream
Fusion~\cite{coutts_stream_2007} for |concatMap|. As the authors and
Coutts say, |concatMap| is complicated because its
mapping function may create any stream whose size is not statically known.
The authors implement Coutts's  idea of
transforming |concatMap| to
|flatten|; the latter supports fusion for a constant
inner stream. Using HERMIT instead of GHC \textsc{Rules}, Farmer et al. present
as motivating examples two cases. Our approach handles the \emph{non-constant
inner stream case} without any additional action.
%
%

The second case is about \emph{multiple inner streams} (of the same state type).
Farmer et al. eliminate some overhead yet do not produce fully
fused code. E.g., pipelines such as the following (in Haskell) are not
fully fused:
\begin{code}
concatMapS (\x -> case even x of
   True -> enumFromToS 1 x
   False -> enumFromToS 1 (x + 1))
\end{code}
(Farmer et al. raise the question of how often such cases arise in a real
program.) Our library internally places no restrictions on inner streams; it
may well be that the flat-mapping function produces streams of different
structure for each element of the outer stream. On the other hand, the
|flat_map| interface only supports nested streams of a fixed
structure---hence with the applicative rather than monadic interface. We can
provide a more general |flat_map| with the continuation-passing
interface for the mapping function, which then implements:
\begin{code}
flat_map_cps (fun x k ->
  .<if (even .~x) then .~(k (enumFromToS ...))
                 else .~(k (enumFromToS ...))>.)
\end{code}
We have refrained from offering this more general interface since there does not seem to be
a practical need.

GHC \textsc{Rules} \cite{jones_playing_2001}, extensively used in Stream
Fusion, are applied to typed code but by themselves are not typed and
are not guaranteed type-preserving. To write GHC rules, one has to have a
very good understanding of GHC optimization passes, to ensure that the
RULE matches and has any effect at all. \textsc{Rules} by themselves offer no
guarantee, even the guarantee that the re-written code is
well-typed. Multi-stage programming ensures
that all staging transformations are type-correct.

Jonnalagedda et al. present a library using only CPS encodings
(fold-based)~\cite{jonnalagedda_fold-based_2015}. It uses the
Gill et al. foldr/build technique~\cite{gill_shortcut_1993} to get staged streams
in Scala. Like foldr/build, it does not support
combinators with multiple inputs such as |zip|.

In our work, we employ the traditional MSP programming model to implement a
performant streaming library. Rompf et al.~\cite{rompf_optimizing_2013}
demonstrate a loop fusion and deforestation algorithm for data parallel loops
and traversals. They use staging as a compiler transformation
pass and apply to query processing for in-memory
objects. That technique lacks the rich range of fused combinators over finite or
infinite sources that we support, but seems adequate for
the case studies presented in that work. Porting our technique from
the staged-library level to the compiler-transformation level may be applicable
in the context of Scala/LMS.

Generalized Stream Fusion~\cite{mainland_generalized_stream_fusion}
puts forward the idea of \emph{bundled} stream representations. Each
representation is designed to fit a
particular stream consumer following the documented cost model.
Although this design does not present a
concrete range of optimizations to fuse combinators and
generate loop-based code directly, it presents a generalized model
that can ``host'' any number of specialized stream
representations. Conceptually, this framework could be used to implement
our optimizations.
However, it relies on the black-box GHC
optimizer---which is the opposite of our approach of full transparency
and portability.

Ziria~\cite{steward_vytiniotis_2015}, a language for wireless
systems' programming, compiles high-level reconfigurable data-flow
programs to vectorized, fused C-code.  Ziria's |tick| and |process|
(pull and push respectively) demonstrate the benefits of having both
processing styles in the same library. It would be interesting to combine
our general-purpose stream library with Ziria's generation of
vectorized C code.

Svensson et al.\cite{svensson_defunctionalizing_2014} unify pull- and
push- arrays into a single library by defunctionalizing push arrays,
concisely explaining why pull and push must co-exist under a unified
library.  They use a compile monad to interpret their embedded
language into an imperative target one. In our work we get that for
free from staging.  Similarly, the representation of arrays in memory, with their
|CMMem| data type, corresponds to staged arrays (of type |'a array code|) in our work.
The library they derive from the defunctionalization of |Push| streams is called
|PushT| and the authors provide evidence that indexing a push array
can, indeed, be efficient (as opposed to simple push-based streams).
The paper does not seem to handle more challenging
combinators like |concatMap| and |take| and does not efficiently
handle the combinations of infinite and finite sources. Still,
we share the same goal: to unify both styles of streams under
one roof. Finally, Svensson et al. target arrays for embedded
languages, while we target arrays natively in the language.  Fusion is
achieved by our library without relying on a compiler to intelligently
handle all corner cases.

\section{Discussion: Why Staging?}
\label{sec:why-staging}

Our approach relies on staging. This may impose a barrier to the
practical use of the library: staging annotations are unfamiliar to
many programmers.  Furthermore, it is natural to ask whether our
approach could be implemented as a compiler optimization pass.


\paragraph{Complexity of staging.}

How much burden staging really imposes on a programmer is an empirical
question. As our library becomes known and more-used we hope to
collect data to answer this. In the meantime, we note that staging can
be effectively hidden in code combinators. The first code example of
\S\ref{sec:overview} (summing the squares of elements of an array) can
be written without the use of staging annotations as:

\begin{lstlisting}
let sum = fold (fun z a -> add a z) zero

of_arr $arr$
  |> map (fun x -> mul x x)
  |> sum
\end{lstlisting}

In this form, the functions that handle stream elements are written
using a small combinator library, with operations |add|, |mul|, etc. that
hide all staging. The operations are defined simply as
\begin{code}
let add x y = .<.~x + .~y>. and mul x y = .<.~x * .~y>.
let zero = .<0>.
\end{code}

Furthermore, our Scala implementation has no explicit staging
annotations, only \sv{Rep} types (which are arguably less
intrusive). For instance, a simple pipeline is shown below:

\begin{lstlisting}[style=Scala,basicstyle={\small\ttfamily},literate=
   {=>}{{$\Rightarrow$}}2]
def test (xs : Rep[Array[Int]]) : Rep[Int] =
  Stream[Int](xs).filter(d => d % 2 == 0).sum
\end{lstlisting}

\paragraph{Staging vs. compiler optimization.}

Our approach can certainly be cast as an optimization pass.  The
current staging formulation is an excellent blueprint for such a
compiler rewrite. However, staging is both less intrusive and more
disciplined---with high-level type safety guarantees---than changing
the compiler. Furthermore, optimization is guaranteed only with full
control of the compiler. Such control is possible in a domain-specific
language, but not in a general-purpose language, such as the ones we
target.  Relying on a general-purpose compiler for library
optimization is slippery.  Although compiler analyses and
transformations are (usually) sound, they are almost never complete: a
compiler generally offers no guarantee that any optimization will be
successfully applied.\footnote{A recent quote by Ben Lippmeier, discussing RePa
  \cite{keller_repa_2010} on Haskell-Cafe, captures well the
  frustrations of advanced library writers: ``The compilation method
  [...] depends on the GHC simplifier acting in a certain way---yet
  there is no specification of exactly what the simplifier should do,
  and no easy way to check that it did what was expected other than
  eyeballing the intermediate code.  We really need a different
  approach to program optimisation [...]  The [current approach] is
  fine for general purpose code optimisation but not `compile by
  transformation' where we really depend on the transformations doing
  what they're supposed
  to.''---\url{http://mail.haskell.org/pipermail/haskell-cafe/2016-July/124324.html}} There are several instances when an innocuous
change to a program makes it much slower. The compiler is a black box,
with the programmer forced into constantly reorganizing the program in
unintuitive ways in order to achieve the desired
performance.



\section{Conclusions}

We have presented the principles and the design of stream libraries
that support the widest set of operations from past libraries
and also permit elimination of the entire abstraction overhead. The
design has been implemented as the \strymonas{} library, for OCaml and for
Scala/JVM. As confirmed experimentally, our library indeed offers the
highest, guaranteed, and portable performance. Underlying the library
is a representation of streams that captures the
essence of iteration in streaming pipelines. It recognizes which
operators drive the iteration, which contribute to filtering
conditions, whether parts of the stream have linearity properties, and
more. This decomposition of the essence of stream iteration is what
allows us to perform very aggressive optimization, via staging,
regardless of the streaming pipeline configuration.

\acks

We thank the anonymous reviewers of both the program committee and the
artifact evaluation committee for their constructive comments. We
gratefully acknowledge funding by the European Research Council under
grant 307334 (\textsc{Spade}).

\clearpage

\balance

\bibliographystyle{abbrvnat}
\bibliography{main}


\clearpage
\newpage
\appendix

\section{Generated code for the Complex example}
\label{app:complex-ex}

We show the generated code for the last example of Section
\S\ref{sec:overview}, repeated below for reference:
\begin{code}
(* Zipping function *)
zip_with (fun e1 e2 -> .<(.~e1,.~e2)>.)
 (* First stream to zip *)
 (of_arr .<arr1>.
   |> map (fun x -> .<.~x * .~x>.)
   |> take .<12>.
   |> filter (fun x -> .<.~x mod 2 = 0>.)
   |> map (fun x -> .<.~x * .~x>.))
 (* Second stream to zip *)
 (iota .<1>.
   |> flat_map (fun x -> iota .<.~x+1>. |> take .<3>.)
   |> filter (fun x -> .<.~x mod 2 = 0>.))
 |> fold (fun z a -> .<.~a :: .~z>.) .<[]>.
\end{code}

The generated code is:
\begin{code}
let s_23 = ref [] in
let arr_24 = arr1 in
 let i_25 = ref 0 in
 let curr_26 = ref None in
 let nadv_27 = ref None in
 let adv_32 () =
   curr_26 := None;
   while
     ((! curr_26) = None) &&
       ((! nadv_27 <> None) ||
          (! i_25 <= (min (12 - 1)
                          (Array.length arr_24 - 1))))
     do
     match ! nadv_27 with
      | Some adv_28 -> adv_28 ()
      | None  ->
          let el_29 = arr_24.(! i_25) in
          let t_30 = el_29 * el_29 in
          incr i_25;
          if (t_30 mod 2) = 0
          then let t_31 = t_30 * t_30 in
               curr_26 := Some t_31
     done in
 adv_32 ();
 let s_33 = ref (Some (1, (1 + 1))) in
 let term1r_34 = ref (! curr_26 <> None) in
 while ! term1r_34 && ! s_33 <> None do
    match ! s_33 with
    | Some (el_35,s'_36) ->
        s_33 := (Some (s'_36, (s'_36 + 1)));
        let s_37 =
           ref (Some (el_35 + 1,
                     (el_35 + 1) + 1)) in
        let nr_38 = ref 3 in
        while (! term1r_34) &&
              (((! nr_38) > 0) &&
               ((! s_37) <> None)) do
          match ! s_37 with
            | Some (el_39,s'_40) ->
                s_37 := Some (s'_40, (s'_40 + 1));
                decr nr_38;
                if el_39 mod 2 = 0
                then
                 (match ! curr_26 with
                   | Some el_41 ->
                     adv_32 ();
                     term1r_34 := !curr_26 <> None;
                     s_23 := (el_41, el_39) :: ! s_23)
            done
    done;
! s_23
\end{code}

\section{Cartesian Product}
\label{App:Cart}

\begin{code}
let cart = fun (arr1, arr2) ->
  ofArr arr1
  |> flat_map (fun x ->
       ofArr arr2 |> map (fun y -> .< .~x * .~y>.))
  |> fold (fun z a -> .<.~z + .~a>.) .<0>.;;
\end{code}

\section{Generated code for Cartesian Product}
\label{App:Cart_code}
\begin{code}
let x = Array.init 1000 (fun i_1  -> i_1) in
let y = Array.init 10   (fun i_2  -> i_2) in
let arr_1 = x in
let size_1 = Array.length arr_1 in
let iarr_1 = ref 0 in
let rec loop_1 acc_1 =
  if (!iarr_1) >= size_1
  then acc_1
  else
    (let el_1 = arr_1.(!iarr_1) in
     incr iarr_117;
     (let acc1_tmp =
        let arr_2 = y in
        let size_2 = Array.length arr_2 in
        let iarr_2 = ref 0 in
        let rec loop_2 acc_2 =
          if (!iarr_2) >= size_2
          then acc_2
          else
            (let el_2 = arr_2.(!iarr_2) in
             incr iarr_2;
             (let acc2_tmp =
                acc_2 + (el_1  * el_2) in
              loop_2 acc2_tmp)) in
        loop_2 acc_1 in
      loop_1 acc1_tmp)) in
loop_1 0
\end{code}

\section{Streams and baseline benchmarks}
\label{App:Bench}
\begin{code}
let sumS
= fun arr ->
   of_arr arr
   |> fold (fun z a -> .<.~z + .~a>.) .<0>.;;

let sumShand
= fun arr1 -> .<
   let sum = ref 0 in
   for counter1 = 0 to Array.length .~arr1 - 1 do
      sum := !sum + (.~arr1).(counter1);
   done;
   !sum >.;;

let sumOfSquaresS
= fun arr ->
   of_arr arr
   |> map (fun x -> .<.~x * .~x>.)
   |> fold (fun z a -> .<.~z + .~a>.) .<0>.;;

let sumOfSquaresShand
= fun arr1 -> .<
   let sum = ref 0 in
   for counter1 = 0 to Array.length .~arr1 - 1 do
    let item1 = (.~arr1).(counter1) in
    sum := !sum + item1*item1;
   done;
   !sum>.;;

let mapsS
= fun arr ->
   of_arr arr
   |> map (fun x -> .<.~x * 1>.)
   |> map (fun x -> .<.~x * 2>.)
   |> map (fun x -> .<.~x * 3>.)
   |> map (fun x -> .<.~x * 4>.)
   |> map (fun x -> .<.~x * 5>.)
   |> map (fun x -> .<.~x * 6>.)
   |> map (fun x -> .<.~x * 7>.)
   |> fold (fun z a -> .<.~z + .~a>.) .<0>.;;

let maps_hand
= fun arr1 -> .<
   let sum = ref 0 in
   for counter1 = 0 to Array.length .~arr1 - 1 do
   let item1 = (.~arr1).(counter1) in
    sum := !sum + item1*1*2*3*4*5*6*7;
   done;
   !sum>.;;

let filtersS
= fun arr ->
   of_arr arr
   |> filter (fun x -> .<.~x > 1>.)
   |> filter (fun x -> .<.~x > 2>.)
   |> filter (fun x -> .<.~x > 3>.)
   |> filter (fun x -> .<.~x > 4>.)
   |> filter (fun x -> .<.~x > 5>.)
   |> filter (fun x -> .<.~x > 6>.)
   |> filter (fun x -> .<.~x > 7>.)
   |> fold (fun z a -> .<.~z + .~a>.) .<0>.;;

let filters_hand
= fun arr1 -> .<
   let sum = ref 0 in
   for counter1 = 0 to Array.length .~arr1 - 1 do
      let item1 = (.~arr1).(counter1) in
      if (item1 > 1 && item1 > 2 && item1 > 3 &&
        item1 > 4 && item1 > 5 && item1 > 6 &&
        item1 > 7) then
      begin
      sum := !sum + item1;
      end;
   done;
   !sum>.;;

let sumOfSquaresEvenS
= fun arr ->
   of_arr arr
   |> filter (fun x -> .<.~x mod 2 = 0>.)
   |> map (fun x -> .<.~x * .~x>.)
   |> fold (fun z a -> .<.~z + .~a>.) .<0>.;;

let sumOfSquaresEvenShand
= fun arr1 -> .<
   let sum = ref 0 in
   for counter1 = 0 to Array.length .~arr1 - 1 do
   let item1 = (.~arr1).(counter1) in
   if item1 mod 2 = 0 then
   begin
     sum := !sum + item1*item1
   end;
   done;
   !sum>.;;

let cartS
= fun (arr1, arr2) ->
   of_arr arr1
   |> flat_map (fun x ->
      of_arr arr2 |> map (fun y -> .< .~x * .~y>.))
   |> fold (fun z a -> .<.~z + .~a>.) .<0>.;;

let cartShand
= fun (arr1, arr2) -> .<
   let sum = ref 0 in
   for counter1 = 0 to Array.length .~arr1 - 1 do
      let item1 = (.~arr1).(counter1) in
      for counter2 = 0 to Array.length .~arr2 - 1 do
         let item2 = (.~arr2).(counter2) in
         sum := !sum + item1 * item2;
      done;
   done;
   !sum >.;;

let dotProductS
= fun (arr1, arr2) ->
   zip_with (fun e1 e2 -> .<.~e1 * .~e2>.)
            (of_arr arr1) (of_arr arr2)
   |> fold (fun z a -> .<.~z + .~a>.) .<0>.;;

let dotProductShand
= fun (arr1, arr2) -> .<
   let sum = ref 0 in
   for counter = 0 to
         min (Array.length .~arr1)
             (Array.length .~arr2) - 1 do
      let item1 = (.~arr1).(counter) in
      let item2 = (.~arr2).(counter) in
      sum := !sum + item1 * item2;
   done;
  !sum>.;;

let flatMap_after_zipWithS
= fun (arr1, arr2) ->
   zip_with (fun e1 e2 -> .<.~e1 + .~e2>.)
            (of_arr arr1) (of_arr arr1)
   |> flat_map (fun x -> of_arr arr2
          |> map (fun el -> .<.~el + .~x>.))
   |> fold (fun z a -> .<.~z + .~a>.) .<0>.;;

let flatMap_after_zipWithShand
= fun (arr1, arr2) -> .<
   let sum = ref 0 in
   for counter1 = 0 to Array.length .~arr1 - 1 do
     let x = (.~arr1).(counter1)
             + (.~arr1).(counter1) in
     for counter2 = 0 to Array.length .~arr2 - 1 do
     let item2 = (.~arr2).(counter2) in
       sum := !sum + item2 + x;
     done;
   done;
   !sum>.;;

let zipWith_after_flatMapS
= fun (arr1, arr2) ->
   of_arr arr1
   |> flat_map (fun x ->
      of_arr arr2 |> map (fun y -> .<.~y + .~x>.))
   |> zip_with (fun e1 e2 -> .<.~e1 + .~e2>.)
               (of_arr arr1)
   |> fold (fun z a -> .<.~z + .~a>.) .<0>.;;

let zipWith_after_flatMapShand
= fun (arr1, arr2) -> .<
   let sum = ref 0 in
   let i1 = ref 0 in
   let i2 = ref 0 in
   let flag1 =
      ref ((!i1) <= ((Array.length .~arr1) - 1)) in
   while (!flag1) &&
         ((!i2) <= ((Array.length .~arr2) - 1)) do
      let el2 = (.~arr2).(!i2) in
      incr i2;
      (let i_zip = ref 0 in
       while (!flag1) &&
             ((!i_zip)
               <= ((Array.length .~arr1) - 1)) do
         let el1 = (.~arr1).(!i_zip) in
         incr i_zip;
         let elz = (.~arr1).(!i1) in
         incr i1;
         flag1 := ((!i1) <=
                   ((Array.length .~arr1) - 1));
         sum := ((!sum) + (elz + el1 + el2))
         done)
      done;
   !sum>.;;

let flat_map_takeS
= fun (arr1, arr2) ->
   of_arr arr1
   |> flat_map (fun x -> of_arr arr2
      |> map (fun y -> .< .~x * .~y>.))
   |> take .<20000000>.
   |> fold (fun z a -> .<.~z + .~a>.) .<0>.;;

let flat_map_takeShand
= fun (arr1, arr2) -> .<
   let counter1 = ref 0 in
   let counter2 = ref 0 in
   let sum = ref 0 in
   let n = ref 0 in
   let flag = ref true in
   let size1 = Array.length .~arr1 in
   let size2 = Array.length .~arr2 in
   while !counter1 < size1 && !flag do
      let item1 = (.~arr1).(!counter1) in
      while !counter2 < size2 && !flag do
         let item2 = (.~arr2).(!counter2) in
         sum := !sum + item1 * item2;
         counter2 := !counter2 + 1;
         n := !n + 1;
         if !n = 20000000 then
         flag := false
      done;
      counter2 := 0;
      counter1 := !counter1 + 1;
   done;
   !sum >.;;
\end{code}

\end{document}